\documentclass[aps,physrev,preprint,groupedaddress]{revtex4-2}
\usepackage{graphicx}
\usepackage{braket}
\usepackage{amsmath}
\usepackage{siunitx}
\usepackage[dvipsnames]{xcolor}
\usepackage{hyperref}
\usepackage{bm}
\newcommand{\dif}{\mathrm{d}}

\DeclareMathOperator{\tr}{tr}

\begin{document}

%\preprint{}

\title{Unified framework of optical thermodynamics and optical pressure}


\author{Nikolaos K. Efremidis}
\email[]{nefrem@uoc.gr}
%\thanks{}
\affiliation{Department of Mathematics and Applied Mathematics, University of Crete, Heraklion 70013, Crete, Greece}
\affiliation{Institute of Applied and Computational Mathematics, Foundation for Research and Technology-Hellas (FORTH), Heraklion 70013, Crete, Greece}

\author{Huizhong Ren}
% \email[]{nefrem@uoc.gr}
%\thanks{}
\affiliation{Ming Hsieh Department of Electrical and Computer Engineering, University of Southern California, Los Angeles 90089, California, USA}

\author{Demetrios N. Christodoulides}
% \email[]{nefrem@uoc.gr}
%\thanks{}
\affiliation{Ming Hsieh Department of Electrical and Computer Engineering, University of Southern California, Los Angeles 90089, California, USA}


\date{\today}

\begin{abstract}
  Optical thermodynamics is a newly developed framework that applies principles from statistical mechanics to describe the intricate behavior of weakly nonlinear, multimode photonic systems. Utilizing this theory, the collective dynamics of complex optical arrangements can be systematically uncovered and understood. The purpose of this work is to examine fundamental aspects of optical thermodynamics, including optical pressure, and provide a unified framework that can be applied to effectively any optical setting within the domain of validity of optical thermodynamics. We find that in addition to the conservation laws, the remaining extensive and intensive parameters of the system are naturally provided by the parameters of the propagation constants. At this point, several thermodynamic approaches exist for analyzing optical forces in multimode settings. Here, we develop a new theoretical methodology that unifies these perspectives in a variety of different configurations, irrespective of whether they are discrete or continuous. We apply our theory in four different settings. By studying Su-Schrieffer-Heeger lattices, we elucidate the thermodynamics of polyatomic chains and show that intercell and intracell bonds can display different optical forces. In addition, we provide a thermodynamic formalism to predict and understand the optical pressure at equilibrium arising in arrangements characterized by a continuous index variation and apply our results to graded-index fibers. 
\end{abstract}

%\keywords{}

\maketitle

\section{Introduction}

Optical thermodynamics aims to describe in a universal manner the statistical response of complex weakly nonlinear multimode photonic systems~\cite{wu-np2019}.
As such it applies to a variety of conservative settings having invariant quantities such as the energy and total power, where different degrees of freedom can be thermalized~\footnote{Systems without such conservation laws require a different description, e.g., open-system or nonequilibrium thermodynamics, and fall outside the scope of the present theory. In addition, the thermodynamics of a photon gas in equilibrium with matter is a different physical problem: photons can be absorbed and emitted by the material reservoir, the photon number $N_\gamma$ is not conserved, and therefore its thermodynamic conjugate variable (the photon chemical potential) vanishes. In this case the probability is described by a canonical ensemble $\rho\sim\exp(-\beta U)$}.
These could include for example, optical multimode and multicore fibers, metallic waveguides, waveguide arrays and photonic lattices, as well as micro-resonators. Central to this theory is the tendency of a system to reach maximum entropy upon equilibration, at which point the modal occupancies follow a Rayleigh-Jeans (RJ) distribution characterized by an optical temperature and chemical potential -- quantities that can be fully predicted from the initial excitation conditions~\cite{parto-ol2019}.
A direct manifestation of optical thermodynamics is the emergence of beam self-cleaning in nonlinear multimode fibers~\cite{lopez-ol2016,liu-ol2016,krupa-np2017}: as the power of the input laser beam increases, the beam profile at the output transitions from a highly speckled pattern to a bell-shaped wavefront. Interestingly, this effect has a statistical origin given that it occurs irrespective of the type of nonlinearity involved~\cite{wu-np2019}. By now, the process of Rayleigh-Jeans thermalization in multimode fibers has been experimentally demonstrated in several studies~\cite{pourb-np2022,mangi-oe2022}. In addition, by modifying the internal energy of the system while keeping the power constant, it is possible to set the Rayleigh-Jeans distribution at a negative optical temperature -- a phenomenon recently observed in time-synthetic lattices~\cite{muniz-science2023} as well as in graded-index multimode fibers~\cite{baudin-prl2023}. The conservation of orbital angular momentum can also be utilized to generalize the resulting Rayleigh-Jeans distribution ~\cite{wu-prl2022,podiv-prl2022}. Joule--Thomson photon-gas expansion processes have recently been observed in photonic lattice configurations~\cite{kirsc-np2025}. On the theoretical front, in addition to fundamental topics of the optical thermodynamic theory~\cite{makri-ol2020,efrem-pra2021}, the process of thermalization from an initial excitation to the final equilibrium distribution has been extensively analyzed~\cite{jung-nc2022,berti-prl2022,ramos-cp2023,yang-aplp2024,yang-aplp2025,ferra-PhysD2025}.

Using energy-based methods, the radiation pressure in photonic structures such as waveguides, couplers, and coupled microresonators has been successfully predicted~\cite{povin-ol2005,povin-oe2005,rakic-oe2009,rakic-ol2011,ren-pra2022} and observed~\cite{li-nature2008,li-np2009,roels-nn2009}. Unlike scattering arrangements, the forces between two evanescently coupled waveguides can be either attractive or repulsive depending on the excitation conditions~\cite{povin-ol2005,povin-oe2005}. A rigorous expression of the optical forces in dielectric structures was derived using the Minkowski-Helmholtz formula~\cite{ren-pra2022}. Such optical forces were observed in experiments involving coupled dielectric waveguides and free-standing lightguides on top of dielectric substrates~\cite{li-nature2008,li-np2009,roels-nn2009}.
A variety of applications of these forces in integrated photonics has been proposed and observed~\cite{vanth-np2010,metac-apr2014}. These range from optical routing~\cite{rosen-np2009} and optical information storage~\cite{huang-acs2019}, to precision measurements~\cite{anets-np2009}, photothermal sensing~\cite{prues-acs2018}, and actuators~\cite{li-np2009,roels-nn2009,ren-ACSNano2013}. Several of these applications exploit the fact that the value of a spatial degree of freedom can change either by optical forces or by external mechanical perturbations. Such displacements, in turn, affect the wave dynamics (for example the power transfer between the waveguides).

Recently, considerable effort has been devoted to investigating the optical forces (due to radiation pressure) in multimode systems at thermal equilibrium~\cite{efrem-cp2022,ren-prl2023}. In this context, it has been demonstrated that by applying the fundamental principles of statistical mechanics, the optical pressure can be calculated thermodynamically, without the need for complex computations based on the Maxwell stress tensor. In~\cite{efrem-cp2022}, using the theory that was developed in~\cite{efrem-pra2021}, a generic formula for the optical pressure was derived, which in the case of photonic lattices can be further simplified [Eqs.~(1) and (2) of Ref.~\cite{efrem-pra2021}, respectively]. In a subsequent work~\cite{ren-prl2023}, the electrodynamic pressure exerted in a nonlinear multimode optical system was explicitly obtained by using purely entropic principles and was found to be in full agreement with Maxwell-stress tensor calculations [Eq.~(9) of Ref.~\cite{ren-prl2023}]. While in Ref.~\cite{efrem-cp2022} a grand-canonical ensemble was utilized, Ref.~\cite{ren-prl2023} was based on a microcanonical ensemble. Interestingly, the formulas in these two works did not seem to be connected. Finally, optical forces have also been examined in composite multimode systems involving several optical species~\cite{ren-ol2024,efrem-ol2024}. Following the discussion of the previous section, we envision that optical forces in highly multimode optical platforms (such as photonic lattices) can be harnessed for reconfigurability and sensing. In the same spirit, small changes of a spatial coordinate (e.g. lattice spacing, boundary), whether induced optically or mechanically, can modify the thermalization properties of the system.

\begin{figure}
\centerline{
\includegraphics[width=\columnwidth]{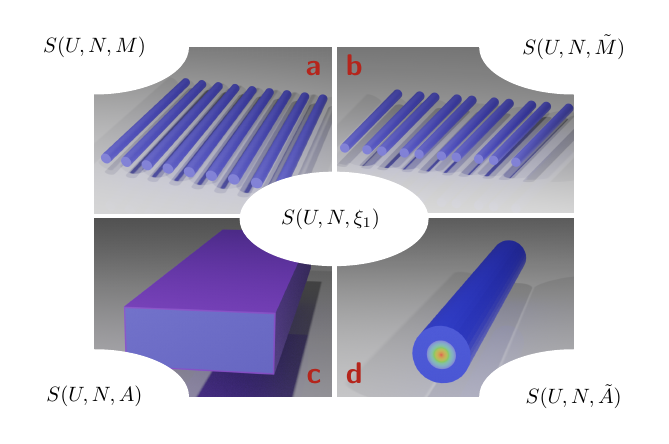}
}
\caption{Discrete and continuous multimode optical platforms used to demonstrate the unified thermodynamic framework. The configurations in the first row are discrete: (a) a regular one-dimensional waveguide array and (b) a diatomic Su-Schrieffer-Heeger lattice, where the intracell and intercell bonds introduce distinct thermodynamic parameters. The second row shows two continuous systems: (c) a metallic waveguide and (d) a graded-index fiber used to investigate settings with non-abrupt boundaries. While the first two extensive natural variables of entropy are always the power $N$ and the internal energy $U$, the third one $\xi_1$ (shown in the middle ellipse) depends on the optical platform, as demonstrated in Sections~\ref{sec:array}-\ref{sec:parabolic}. Specifically, in (a) it is the number of modes $M$, in (b) the number of primitive cells $\tilde M$, in (c) the waveguide length $L$ or area $A$ (depending on dimensionality), and in (d) the effective beam area $\tilde A$.}
\label{fig:settings}
\end{figure}

\begin{figure}
\centerline{
\includegraphics[width=\columnwidth]{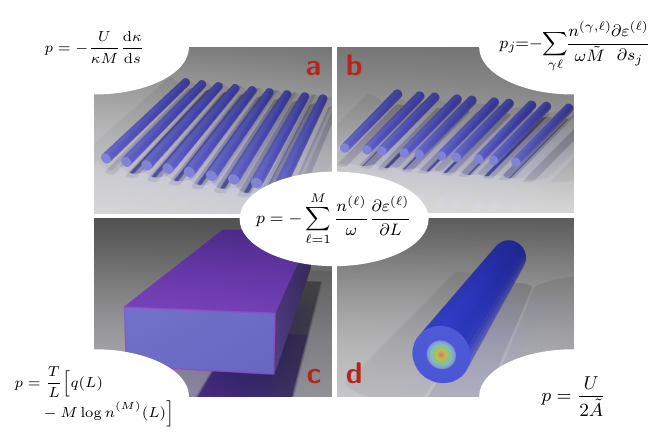}
}
\caption{
    Platform-dependent expressions for optical pressure derived from the unified thermodynamic formalism. The central modal-sum expression [see Eq.~(\ref{eq:FinalPressure})] represents the generic structure of the thermodynamic pressure, from which the platform-specific formulas are derived. Here $L$ denotes the relevant size coordinate of the system, and the derivative is taken for a fixed number of modes. (a) For a regular one-dimensional lattice the summation reduces to the closed-form expression of Eq.~(\ref{eq:pArraySimplified}) for the mean pressure between adjacent waveguides. (b) In a diatomic SSH lattice, two pressures $p_j$, $j=1,2$, are obtained, corresponding to the mean forces on the intracell and intercell bonds [Eq.~(\ref{eq:pjSSH})]. (c) For a bulk metallic waveguide, an asymptotic evaluation of the modal sum leads to Eq.~(\ref{eq:papprox}) for the pressure. This expression agrees with the exact summation formula except near the high-energy condensation limit $T\rightarrow0^-$. (d) For systems with a radial parabolic index profile, such as graded-index fibers, the summation leads to the closed-form expression of Eq.~(\ref{eq:PressureGRIN2Dv2}). Thus, the pressure is proportional to the internal energy and inversely proportional to the effective area of the beam. 
  }
\label{fig:PressureResults}
\end{figure}

In this work, we explore several fundamental aspects that are central to optical thermodynamics. We first revisit the very concept of entropy extensivity. As previous studies have shown, the entropy happens to be extensive with respect to the power $N$, the internal energy (or momentum flow) $U$, and the number of modes $M$. Here, we show that this assertion can be generalized. While $U$ and $N$ are always present in the fundamental thermodynamic equation, the third quantity is a system parameter that affects the spectrum or the propagation constants.
For the systems that we investigate, the third extensive natural variable of entropy can be either the number of modes $M$, the number of primitive cells $\tilde M$, the size of the system $\Omega$, or the effective size of the system $\tilde\Omega$, as summarized schematically in Fig.~\ref{fig:settings}.
Unlike previous approaches~\cite{efrem-cp2022,ren-prl2023}, in this work, the eigenvalues or the propagation constants are directly employed to determine the complete set of parameters (including the third extensive natural variable of entropy) required for analyzing the thermodynamic relations at equilibrium as well as the electrodynamic forces. In this regard, we provide a generalized and unified framework that can be applied to any type of conservative discrete or continuous multimode weakly nonlinear optical settings having invariant quantities such as the energy and total power [see Fig.~\ref{fig:settings}]. In calculating the optical pressure, we compare closed-form results with asymptotic expressions in order to identify in which regime they are valid.
Figure~\ref{fig:PressureResults} summarizes the resulting optical-pressure expressions for the settings considered in this work.
For example, Eq.~(9) of Ref.~\cite{ren-prl2023} for continuous systems is derived using asymptotic calculations from the exact expression for the optical pressure and is found to be very accurate except in the high-energy condensation limit $T\rightarrow0^-$. Thermodynamic pressure effects are also investigated in diatomic lattices such as a Su-Schrieffer-Heeger chain. We unveil that such systems are associated with two different optical pressures related to the two bonds. In such polyatomic settings, the third extensive natural variable of entropy is the number of primitive cells.

We also study systems with continuous index variations. Due to the presence of smooth index changes, the thermal equilibrium beam size is, in general, not characterized by the size of the waveguide. We focus on the particular example of parabolic index waveguides in both one and two transverse directions (graded-index fibers).
We compute compact closed-form expressions for the thermal equilibrium beam size and optical pressure, both of which depend on the waveguide parameters as well as the beam power and internal energy.
We also discuss limitations that arise from the asymptotic character of the Rayleigh-Jeans distribution. In particular, the Rayleigh-Jeans distribution predicts non-zero power occupation numbers even at cutoff, i.e., at the transition between a propagating mode and an evanescent mode or between a bound mode and a radiation mode. As a result, thermodynamic parameters (such as the entropy, the optical temperature, and the chemical potential) might exhibit discontinuities exactly at the cutoff. Prominent examples are the predicted pressure singularities and the, on average, decrease of the entropy in the case of metallic waveguides.

The paper is organized as follows: In Section~\ref{sec:general} we develop the general framework of our results. Then, in Sections~\ref{sec:array}-\ref{sec:parabolic} we apply our theory to four representative systems: one-dimensional lattices, Su-Schrieffer-Heeger lattices, metallic waveguides, and graded-index fibers. Finally, in Section~\ref{sec:limitations} we discuss limitations of the Rayleigh-Jeans distribution as compared to the expected behavior.

\section{General formalism\label{sec:general}}

We consider a weakly nonlinear multimode optical system with a Hamiltonian $\hat H$, and expand the optical wave as
\begin{equation}
  |\Psi\rangle = \sum_{\ell=1}^MC^{(\ell)}(z)|\ell\rangle, 
  \label{eq:Psiexpansion}
\end{equation}
where $|\ell\rangle$ are the supermodes with propagation constants $\varepsilon^{(\ell)}$,  and $C^{(\ell)}(z)$ are the modal amplitudes that depend on the propagation distance $z$.
The type of the nonlinearity (such as Kerr, saturable, or quadratic) is not particularly important, as long as the system is not close to being integrable. 
The system has two conservation laws, namely the total power
\begin{equation}
  N = \langle\Psi|\Psi\rangle =
  \sum_{\ell=1}^M|C^{(\ell)}|^2=
  \sum_{\ell=1}^Mn^{(\ell)}
  \label{eq:N}
\end{equation}
(where $n^{(\ell)}$ is the power of mode $\ell$ or the power occupation number),
and the internal energy (which physically represents the electromagnetic momentum flow, see for example Ref.~\cite{haus-josa1976})
\begin{equation}
  U =
  \frac1\omega
  \langle\Psi|\hat H|\Psi\rangle
  =
  \sum_{\ell=1}^M\frac{\varepsilon^{(\ell)}n^{(\ell)}}{\omega},
  \label{eq:U}
\end{equation}
where $\omega$ is the frequency of the optical wave. Due to the relatively weak strength of the nonlinear effects, in the last expression of Eq.~(\ref{eq:U}) we have approximated the Hamiltonian with its linear part.
Each physical setting has its own set of \textit{system parameters} $\xi=\{\xi_1,\xi_2,\ldots,\xi_J\}$ that enter the propagation constants  $\varepsilon^{(\ell)}(\xi)$ and, in turn, determine the final thermodynamic expressions. Thus, in optical thermodynamics the final relations are system dependent.
We assume that entropy is \textit{extensive} (homogeneous of degree one) with respect to three parameters, two of which are the conservation laws $N$ and $U$, while the last one is the first system parameter $\xi_1$. The parameters $(N,U,\xi_1)$ are called the \textit{extensive natural variables of entropy}. 
We utilize a grand canonical approach by maximizing the Gibbs entropy $S=-\int\rho\log\rho\dif\Gamma$ under the constraints that the total probability is equal to one $\int\rho\dif\Gamma=1$ subject to the mean values
$\braket{N}=\int_\ell\rho N\dif\Gamma$  and
$\braket{U}=\int \rho U\dif\Gamma$, where we integrate over the phase-space $\dif\Gamma=(i/2)^M\prod_{\ell=1}^M\dif C^{(\ell)}\dif C^{(\ell)*}$.
We derive the following probability
\[
  \rho = \exp[
  -q-\alpha N-\beta U
  ]
\]
of the system being associated with power occupation numbers $\{n^{(\ell)}\}$~\cite{pathr-2011}.
In the above expression, $\alpha$ and $\beta$ are the Lagrange multipliers. The $q$-potential is the probability normalization coefficient that is related to the grand canonical partition function $\mathcal Q$ via $q = \log\mathcal Q$. Following the calculations we derive
\begin{equation}
  q =
  \log \int e^{-\alpha N-\beta U}
  \dif\Gamma
  =
  M\log\pi
  -\sum_{\ell=1}^M\log(\alpha+\beta\varepsilon^{(\ell)}/\omega).
  \label{eq:q}
\end{equation}
Differentiating the $q$-potential with respect to $\varepsilon^{(\ell)}$, we find that the ensemble average power occupation numbers follow a Rayleigh-Jeans distribution
\begin{equation}
  \langle n^{(\ell)}\rangle=
  -
  \frac\omega\beta\left(
    \frac{\partial q}{\partial\varepsilon^{(\ell)}}
  \right)_{\alpha,\beta,\overline\varepsilon^{(\ell)}}
  =
  \frac1{\alpha+\beta\varepsilon^{(\ell)}/\omega},
  \label{eq:nl}
\end{equation}
where $\varepsilon=\{\varepsilon^{(1)},\ldots,\varepsilon^{(M)}\}$ and $\overline\varepsilon^{(\ell)}=\varepsilon\setminus\varepsilon^{(\ell)}$. 
In addition, we derive the ensemble average values of the two conservation laws 
\begin{equation}
  \langle N\rangle =
  -
  \left(
    \frac{\partial q}{\partial\alpha}
  \right)_{\beta,\xi}
  = \sum_{\ell=1}^M\langle n^{(\ell)}\rangle,
  \label{eq:N2}
\end{equation}
\begin{equation}
  \langle U\rangle =
  -
  \left(
    \frac{\partial q}{\partial\beta}
  \right)_{\alpha,\xi}
  =
  \frac1\omega
  \sum_{\ell=1}^M\langle n^{(\ell)}\rangle \varepsilon^{(\ell)}.
  \label{eq:U2}
\end{equation}
From this point on, for simplicity, we will omit the brackets that denote ensemble average. By direct substitution, we find the following energy expression
\begin{equation}
  \alpha N + \beta U = M
  \label{eq:energyrelation}
\end{equation} 
that relates the energy, the power, and the number of modes, with the Lagrange multipliers. Taking the differential of $q(\alpha,\beta,\varepsilon(\xi))$, and utilizing Eqs.~(\ref{eq:N2})-(\ref{eq:U2}) we obtain
\begin{equation}
  \dif(q+M) =
  \beta\left[
    \frac\alpha\beta\dif N +\dif U +\sum_{j=1}^JR_j\dif\xi_j
  \right]
  \label{eq:qplusM}
\end{equation}
where
\begin{equation}
  R_j
  =
  T
  \left(
    \frac{\partial q}{\partial\xi_j}
  \right)_{\alpha,\beta,\overline\xi_j}
  =
  -\sum_{\ell=1}^M\frac{n^{(\ell)}}{\omega}
  \left(
    \frac{\partial\varepsilon^{(\ell)}}{\partial\xi_j}
  \right)_{\alpha,\beta,\overline\xi_j}
  \label{eq:Rj}
\end{equation}
is the conjugate variable to $\xi_j$ and $\overline\xi_j=\xi\setminus\xi_j$. Comparing Eq.~(\ref{eq:qplusM}) to the expected differential form $\dif U = T\dif S +\mu\dif N +\cdots$, we find that the Lagrange multipliers are related to the optical temperature $T$ and the chemical potential $\mu$ via $\beta=1/T$, $\alpha=-\mu/T$, and the entropy is proportional to the $q$-potential
\[
  S = q+M. 
\]
Thus, the Rayleigh-Jeans distribution of Eq.~(\ref{eq:nl}) takes the form
\begin{equation}
  n^{(\ell)}=\frac{T}{\varepsilon^{(\ell)}/\omega-\mu},
  \label{eq:nl2}
\end{equation}
the energy relation given by Eq.~(\ref{eq:energyrelation}) becomes
\begin{equation}
  U-\mu N = MT,
  \label{eq:energyrelation2}
\end{equation}
and the differential of the internal energy can be written as
\begin{equation}
  \dif U = T\dif S +\mu\dif N - \sum_{j=1}^JR_j\dif\xi_j.
  \label{eq:difU}
\end{equation}
Since the power occupation numbers $n^{(\ell)}$ are positive numbers, from Eq.~(\ref{eq:nl2}) we directly see that for positive temperatures, lower modes have higher occupation numbers. On the other hand, for negative temperatures the higher modes have larger power occupancies.

We assume that the entropy is extensive with respect to $U,N,\xi_1$ and \textit{intensive} (homogeneous of degree zero) with respect to $\xi_2,\ldots,\xi_J$
\begin{equation}
  S(\lambda U,\lambda N,\lambda\xi_1;\xi_2,\ldots,\xi_J)=
  \lambda S(U,N,\xi_1;\xi_2,\ldots,\xi_J).
  \label{eq:entropy_ext}
\end{equation}

Thus, we can proportionally integrate the extensive variables in Eq.~(\ref{eq:difU}) while keeping the intensive variables constant leading to 
\begin{equation}
  U = TS +\mu N - R_1\xi_1,
  \label{eq:U3}
\end{equation}
an expression that can also be derived through Euler's homogeneous function theorem. 
Combining Eqs.~(\ref{eq:difU})-(\ref{eq:U3}) we obtain a Gibbs-Duhem relation
\[
  S\dif T + N \dif\mu-\xi_1\dif R_1+\sum_{j=2}^JR_j\dif\xi_j=0
\]
that relates the variations between all the intensive parameters. 
Finally, from Eq.~(\ref{eq:energyrelation}) and (\ref{eq:U3}) we find that $R_1$ can be written as
\begin{equation}
  R_1=\frac{Tq}{\xi_1}.
  \label{eq:R1}
\end{equation}
In the next sections, we will apply this general formalism to different optical settings.

\section{One-dimensional lattice \label{sec:array}}

Let us first utilize the results of Section~\ref{sec:general} to the case of a regular one-dimensional photonic lattice with either zero or periodic boundary conditions [see for example Fig.~\ref{fig:settings}(a)]. The statistical mechanics and the optical pressure of this system are examined in detail in~\cite{efrem-cp2022}.
As we discussed above, the additional system parameters $\xi_j$ are directly derived from the propagation constants.
For single-mode waveguides the propagation constants are given by
\begin{equation}
  \varepsilon^{(\ell)}(M,s)=-2\kappa(s)
  \cos\frac{\pi(1+\zeta)\ell}{M+1-\zeta},\quad \ell=1,\ldots,M,
  \label{eq:epsilon1Darray}
\end{equation}
where $\zeta=0$ for homogeneous Dirichlet boundary conditions (or open boundaries) and $\zeta=1$ for periodic boundary conditions. In the above expressions $M$ is the number of waveguides (and thus the number of modes), and $\kappa$ is the coupling coefficient that depends on the spacing $s$ between two adjacent waveguides. Thus, the third extensive natural variable of entropy is $\xi_1=M$ whereas $\xi_2=s$ is intensive. The  variable conjugate to the number of modes which we call ``internal pressure'', is given by
\begin{equation}
  R =
  T
  \left(
    \frac{\partial q}{\partial M}
  \right)_{\alpha,\beta,s}
  =
  -\sum_{\ell=1}^M
  \frac{n^{(\ell)}}{\omega}
  \left(
    \frac{\partial\varepsilon^{(\ell)}}{\partial M}
  \right),
  \label{eq:R}
\end{equation}
and characterizes how the internal energy will change when we increase the number of modes. The actual average value of the optical pressure between waveguides is the intensive variable which is conjugate to the size of the system $L\approx sM$. Specifically, using Eq.~(\ref{eq:Rj}) with $\xi_2=s$ and dividing by $M$, we find that the optical pressure $p$ is given by 
\[
  p = -\sum_{\ell=1}^M\frac{n^{(\ell)}}{\omega M}
  \left(
    \frac{\partial\varepsilon^{(\ell)}}{\partial s}
  \right)_{\alpha,\beta,M}.
\]
The above expression can be further simplified by utilizing Eq.~(\ref{eq:epsilon1Darray}) for the propagation constants
\begin{equation}
  p = -\frac{U}{M}\frac{\dif\log\kappa}{\dif s}.
  \label{eq:pArraySimplified}
\end{equation}
We observe that the pressure is directly proportional to the internal energy per mode ($U/M$) and the derivative of $\log\kappa$ with respect to the spacing.
The thermodynamic pressure in such uniform waveguide arrays represents the average value of the optical forces per unit of propagation length exerted between adjacent waveguides. The total applied force per unit of propagation length in each waveguide is the vectorial sum of the forces due to its first neighbor(s) (see Ref.~\cite{efrem-cp2022} for details).
From Eq.~(\ref{eq:difU}) the differential of the internal energy becomes
\[
  \dif U = T\dif S + \mu\dif N -pM\dif s -R\dif M. 
\]

If we further assume extensivity, then we can directly integrate $\dif U$ leading to
\begin{equation}
  U = TS+\mu N -RM.
  \label{eq:latextU}
\end{equation}
In addition, we  derive a Gibbs-Duhem equation
\begin{equation}
  S\dif T +N\dif\mu+pM\dif s-M\dif R = 0
  \label{eq:latextGD}
\end{equation}
as well as a simplified formula for the internal pressure \begin{equation}
  R=Tq/M.
  \label{eq:latextR}
\end{equation}

\section{Su-Schrieffer-Heeger lattice \label{sec:SSH}}

Polyatomic photonic lattices can, in principle, exhibit a more complex behavior due to the presence of additional system parameters. In particular, here we focus on the thermodynamics of a one-dimensional Su-Schrieffer-Heeger (SSH) lattice with alternating coupling coefficients [Fig.~\ref{fig:settings}(b)]. In the case of a topologically trivial configuration, the propagation constants are arranged in two bands according to
\begin{equation}
  \varepsilon(k^{(\ell)}(\tilde M),s_1,s_2;\gamma)
  =
  (-1)^\gamma\sqrt{\kappa_1^2+\kappa_2^2+2\kappa_1\kappa_2\cos k^{(\ell)}},
  \label{eq:epsilonSSH}
\end{equation}
where $\kappa_1(s_1)$ is the intracell and $\kappa_2(s_2)$ is the intercell coupling coefficient, which depend on the two different spacings between the waveguides ($s_1$ and $s_2$, respectively), and $\gamma=1,2$ is the band index. Thus, the length of the primitive cell is $s=s_1+s_2$. When the boundary conditions are periodic, the Bloch momentum is given by $k^{(\ell)}(\tilde M)=2\pi\ell/\tilde M$, where $\ell=1,\ldots,\tilde M$, and $\tilde M=M/2$ is the number of primitive cells. On the other hand, for homogeneous Dirichlet boundary conditions in the topologically trivial regime $\kappa_2/\kappa_1<1+1/\tilde M$, the transverse wavenumbers are derived by solving a transcendental equation~\cite{efrem-pra2021ssh}. 

From Eq.~(\ref{eq:epsilonSSH}), we see that besides $\tilde M$, the propagation constants are also functions of the two different spacings $s_1$ and $s_2$. The conjugate variable to $\tilde M$ is the internal pressure defined by Eq.~(\ref{eq:R}). Since $s_1$, $s_2$ represent lengths, their conjugate variables are extensive $\tilde M p_1$ and $\tilde M p_2$, where $p_1,p_2$ are the thermodynamic pressures exerted between adjacent waveguides due to the intercell and intracell couplings 
\begin{equation}
  p_j = -
  \sum_{\gamma=1}^2\sum_{\ell=1}^{\tilde M}
  \frac{n^{(\gamma,\ell)}}{\omega\tilde M}
  \left(
    \frac{\partial\varepsilon^{(\gamma,\ell)}}{\partial s_j}
  \right)_{\alpha,\beta,\tilde M,s_{3-j}},\quad j=1,2.
  \label{eq:pjSSH}
\end{equation}
These pressures represent the average values of the optical forces per unit of propagation length and are applied to the waveguides along directions defined by the lines connecting the centers of neighboring waveguides. 
The differential of the internal energy contains these two distinct pressure terms
\[
  \dif U = T\dif S+\mu\dif N-R\dif \tilde M
  -p_1\tilde M \dif s_1  -p_2\tilde M \dif s_2.
\]
Alternatively, the differential of the internal energy
\begin{equation}
  \dif U = T\dif S+\mu\dif N-R\dif \tilde M
  -p\tilde M \dif s  -\tau\tilde M\dif r.
  \label{eq:difU:SSH2}
\end{equation}
can be expressed in terms of the average pressure
\[
  p = \frac{p_1+p_2}{2}
  =
  -\sum_{\gamma=1}^2\sum_{\ell=1}^{\tilde M}
  \frac{n^{(\gamma,\ell)}}{\omega\tilde M}
  \left(
    \frac{\partial\varepsilon^{(\gamma,\ell)}}{\partial s}
  \right)_{T,\mu,\tilde M,r}
\]
and the pressure differential between the two bonds
\begin{equation}
  \tau = \frac{p_1-p_2}2
  =
  -\sum_{\gamma=1}^2\sum_{\ell=1}^{\tilde M}
  \frac{n^{(\gamma,\ell)}}{\omega\tilde M}
  \left(
    \frac{\partial\varepsilon^{(\gamma,\ell)}}{\partial r}
  \right)_{T,\mu,\tilde M,s},
  \label{eq:strain}
\end{equation}
where $r=s_1-s_2$ is the intercell-intracell length difference. 

Assuming that the system is extensive, we can derive a Gibbs-Duhem relation
\[
  S\dif T + N\dif\mu
  +p_1\tilde M\dif s_1
  +p_2\tilde M\dif s_2 
  -\tilde M\dif R
  = 0 
\]
which can also be written in terms of $s$, $r$ as
\[
  S\dif T + N\dif\mu
  +p\tilde M\dif s
  +\tau\tilde M\dif r 
  -\tilde M\dif R
  = 0 .
\]
Finally, Eqs.~(\ref{eq:latextU}), (\ref{eq:latextR}) have the same form as compared to the case of a monoatomic lattice (with $M\rightarrow\tilde M$).

Lattice settings, such as those studied in Sections~\ref{sec:array}-\ref{sec:SSH}, might be experimentally observed and investigated for potential applications. The main requirement is the use of a suspended array to magnify the mechanical effects of the optical forces. For example, in Ref.~\cite{efrem-cp2022}, we have studied an array suspended in an index matching fluid. Our calculations showed that according to the optical temperature (that depends on the beam excitation condition), the optical pressure (force per unit of propagation length) can take values in the range $-0.587\ \unit{\nano\newton\per\micro\meter}\le p\le 0.587 \unit{\nano\newton\per\micro\meter}$.

\section{Metallic waveguides\label{sec:metal}}

We now proceed to the investigation of continuous systems. In this section we examine the case of a one-dimensional metallic waveguide [Fig.~\ref{fig:settings}(c) with one direction being significantly longer than the other]. The effect of thermodynamic pressure in such waveguides is examined in detail in~\cite{ren-prl2023}. The modes of the waveguide can be either transverse electric or transverse magnetic having in both cases propagation constants
\begin{equation}
  \varepsilon^{(\ell)}(L) = \sqrt{k^2-\frac{\ell^2\pi^2}{L^2}}. 
  \label{eq:epsilonmetallic}
\end{equation}
The number of propagating modes can be found by requiring the expression inside the square root to be positive
\[
  M =
  \left\lfloor
  \frac{Lk}{\pi}
  \right\rfloor.
\]
From Eq.~(\ref{eq:epsilonmetallic}), we see that a main difference, as compared to the discrete configurations studied in Sections~\ref{sec:array}-\ref{sec:SSH}, is that the propagation constants do not depend on the number of modes $M$ but rather on the size of the waveguide $L$. Thus, we should select $L$ as the third extensive natural variable of entropy. The pressure is the conjugate variable of $L=\xi_1$ which, utilizing Eq.~(\ref{eq:Rj}), is given by
\begin{equation}
  p = T
  \left(
    \frac{\partial q}{\partial L}
  \right)_{\alpha,\beta}
  =
  -\sum_{\ell=1}^M
  \frac{n^{(\ell)}}{\omega}
  \frac{\dif\varepsilon^{(\ell)}}{\dif L}.
  \label{eq:PressMetal0}
\end{equation}
Substituting the propagation constants [Eq.~(\ref{eq:epsilonmetallic})], we can express the thermodynamic pressure as
\begin{equation}
  p = -\sum_{\ell=1}^M
  \frac{n^{(\ell)}}{\omega}
  \frac{\pi^2\ell^2}{L^3\varepsilon^{(\ell)}}.
  \label{eq:PressMetal}
\end{equation}
Equation~(\ref{eq:PressMetal}) represents the mean optical force per unit of propagation length that is applied on the waveguide. From Eq.~(\ref{eq:difU}) the differential of the internal energy takes the form
\begin{equation}
  \dif U = T\dif S +\mu\dif N -p\dif L.
  \label{eq:difUmetal}
\end{equation}
Assuming that the system is extensive, we can directly integrate the extensive variables while keeping the intensive variables constant. Thus from Eq.~(\ref{eq:difUmetal}) we derive the following relation for the internal energy 
\begin{equation}
  U = TS + \mu N - pL.
  \label{eq:Umetal}
\end{equation}
Combining Eq.~(\ref{eq:Umetal}) with the energy relation of Eq.~(\ref{eq:energyrelation2}), we find the following simplified expression for the pressure
\begin{equation}
  p = \frac{Tq}{L}. 
  \label{eq:PressMetalApprox}
\end{equation}
An alternative way to derive Eq.~(\ref{eq:PressMetalApprox}) is by starting from the first equality of Eq.~(\ref{eq:PressMetal0}), and noting that the variables $\alpha$ and $\beta$ (or equivalently $T,\mu$) that are kept constant are intensive. Since both $S$ and $M$ extensively increase with $(N,U,L)$, we conclude that $q=S-M$ is also extensive. Thus, we can approximate $(\partial q/\partial L)_{T,\mu}\approx q/L$ leading to Eq.~(\ref{eq:PressMetalApprox}).

With the above reasoning, since $q$ is extensive, we can on average approximate its partial derivative $(\partial q/\partial L)_{T,\mu}$ with $q/L$. However, if the function $q$ has discontinuities, then the average slope can, in general, be quite different from the local slope. In our numerical results we find that such discontinuities exist, and are located at the values $L=M\pi/k$ where additional modes are introduced. Since the value of the pressure is dictated from the local slope, rather than the average slope, Eq.~(\ref{eq:PressMetalApprox}) fails to accurately approximate the pressure.

Our task is to find a way to approximate the average local value of the partial derivative appearing in Eq.~(\ref{eq:PressMetal0}). We define the function $F(m)$ to be the waveguide length at which the $m$th mode is introduced. In the case of metallic waveguides $F(m)=m\pi/k$.  Thus, in the right limit $F(m)^+ = \lim_{\epsilon\rightarrow0^+}(F(m)+\epsilon)=m\pi/k+\lim_{\epsilon\rightarrow0^+}\epsilon$ the system will have $m$ modes, and in the left limit $F(m)^- = \lim_{\epsilon\rightarrow0^-}(F(m)+\epsilon)=m\pi/k+\lim_{\epsilon\rightarrow0^-}\epsilon$ the system will have $m-1$ modes. We can express the discontinuity in the $q$-potential due to the $m$th mode as
\begin{equation}
  \Delta q_m =
  q(F(m)^+)-q(F(m)^-)
  = 
  \sum_{\ell=1}^{m-1}\log
  \frac{n^{(\ell)}(F(m)^+)}{n^{(\ell)}(F(m)^-)}
  +\log n^{(m)}(F(m)^+).
  \label{eq:dqMMetal}
\end{equation}
Note that ideally, asymptotically close to the cutoff, the left and right limits of the power occupation numbers should be identical, meaning that $n^{(\ell)}(F(m)^+)=n^{(\ell)}(F(m)^-)$ for $\ell<m$ and $n^{(m)}(F(m)^+)=0$. If these conditions are satisfied, then the discontinuity in the $q$-potential becomes zero $\Delta q_m=0$ and the asymptotic expression for the pressure given by Eq.~(\ref{eq:PressMetalApprox}) should be valid. However, due to the asymptotic nature of the Rayleigh-Jeans distribution, we observe a non-zero power occupation number of the last mode at the cutoff $n^{(m)}(F(m)^+)>0$ and thus, since the total power is constant [($N,U,L$) increase proportionally], in general $n^{(\ell)}(F(m)^+)\neq n^{(\ell)}(F(m)^-)$ for $\ell<m$. 

To circumvent this problem and obtain an approximate average value for the partial derivative of $q$ with respect to $L$, we can subtract from Eq.~(\ref{eq:PressMetalApprox}) all the discontinuities due to the introduction of the $1$st, $2$nd, $\ldots$, $M$th mode, leading to
\begin{equation}
  \left(
    \frac{\partial q}{\partial L}
  \right)_{\alpha,\beta}
  \approx
  \frac1L
    \left[
      q(L)-\sum_{m=1}^M\Delta q_m
    \right].
\end{equation}
The above expression requires the calculation of all the power occupation numbers at all the discontinuities. Ideally, we would like to find a simpler expression. Following our original assumption that the system is extensive then, as the number of modes increases, all the discontinuities $\Delta q_m$ will approximately have the same value. Thus, we can keep only the last discontinuity $m=M$ multiplied by the number of discontinuities
\begin{equation}
  \left(
    \frac{\partial q}{\partial L}
  \right)_{\alpha,\beta}
  \approx
  \frac{q(L)-M\Delta q_M}L.
  \label{eq:PressMetalApprox2}
\end{equation}
Since we assume that $M\gg1$, the introduction of the last mode is not going to significantly change the values of the existing power occupation numbers $n^{(\ell)}$. This happens because only a small amount of power will be transferred from each one of the existing modes $\ell=1,\ldots,M-1$ to the last mode $\ell=M$. Thus, the terms appearing in the sum on the right hand side of Eq.~(\ref{eq:dqMMetal}) will be approximately zero, with the main contribution coming from the last term. We conclude that
\begin{equation}
  \Delta q_M \approx
  \log n^{(M)}(F(M)^+),
  \label{eq:dqMMetal2}
\end{equation}
a formula that depends only on a single power occupation number. 
We expect that the above approximations will start to fail as, under constant power, the internal energy increases, and especially when we reach negative optical temperatures. In this regime, the power distribution $n^{(\ell)}$ is inverted and the highest mode has the largest occupancy. As a result the fractions $n^{(\ell)}(F(\ell)^+)/n^{(\ell)}(F(\ell)^{-})$ will have significant deviations from unity, and the terms appearing in the sum in Eq.~(\ref{eq:dqMMetal}) cannot be ignored.

We conclude that the optical pressure can be approximated by
\begin{equation}
  p\approx
  \frac TL
  [q(L)-M\log n^{(M)}(L)]
  \label{eq:papprox}
\end{equation}
Equation~(\ref{eq:papprox}) was originally derived in Ref.~\cite{ren-prl2023} using a different approach. As we showed above, it is expected to be valid in the lower energy regime and, in particular, when $T>0$. For negative temperatures, it starts to fail especially as we approach the high energy condensation limit $T\rightarrow0^-$.

%\section{Figure 1}
\begin{figure}
\centerline{
\includegraphics[width=\columnwidth]{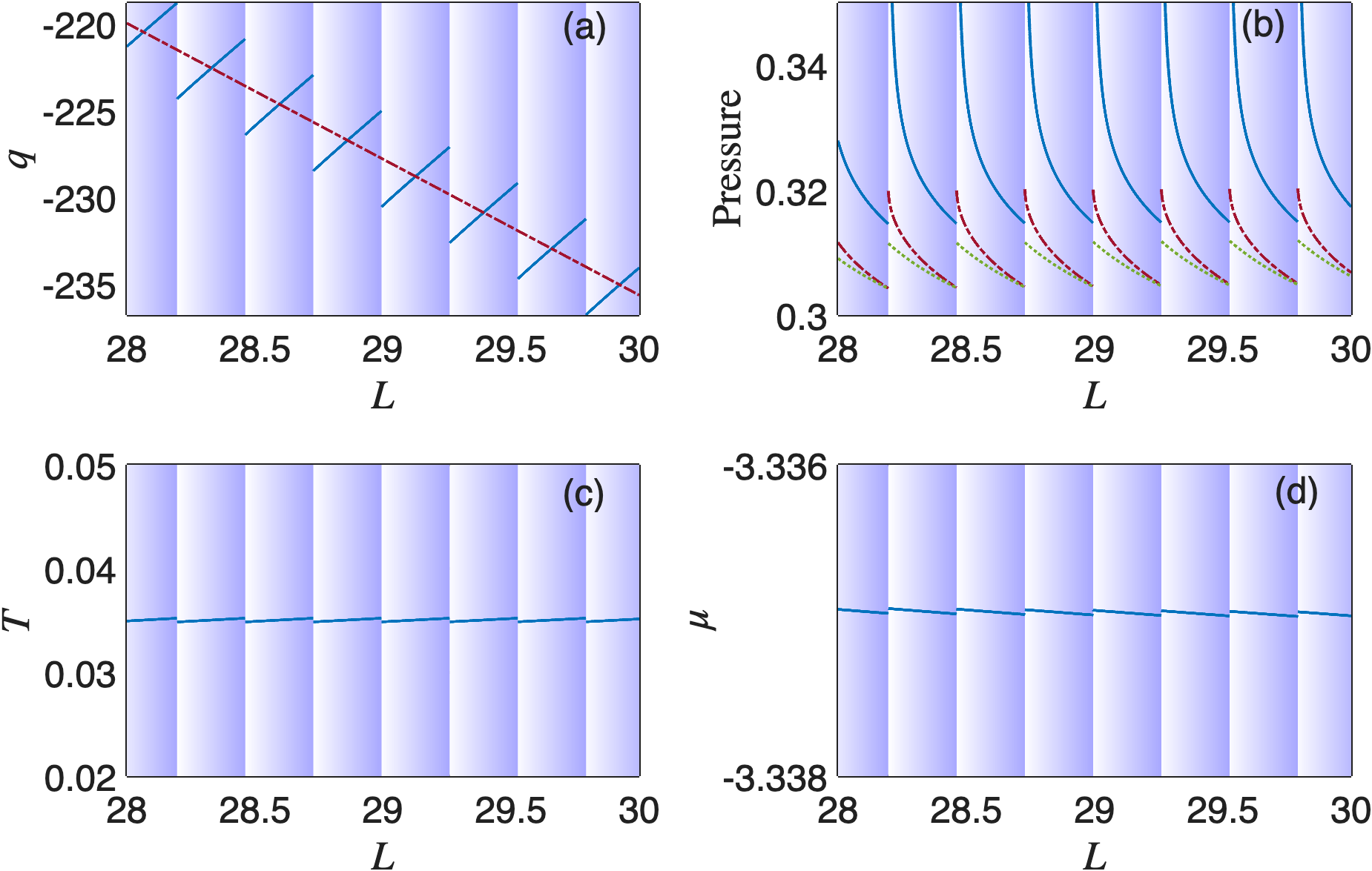}
}
\caption{Proportional variation of the set of extensive natural variables of entropy ($L, N, U$) for a metallic waveguide. To display the effect of the introduction of new modes, on the horizontal axis we vary $L$ for a relatively small range of values. (a) The $q$-potential shown with blue lines is piecewise continuous with discontinuities at the locations where new modes are introduced. It is extensive and thus its average value can be described by the red dashed-dotted line that passes through the origin. (b) Exact optical pressure using Eq.~(\ref{eq:PressMetal}) is shown with blue lines, an approximation where the last mode of the sum is truncated is shown with green dotted lines, whereas the red dashed-dotted lines depict the approximate calculation for the pressure using Eq.~(\ref{eq:papprox}). In (c) and (d), we present the temperature and the chemical potential which, due to their intensive character, remain almost invariant as a function of $L$.}
\label{fig:1}
\end{figure}
In Fig.~\ref{fig:1}, we proportionally increase all the extensive natural variables of entropy of the metallic waveguide $(L,N,U)$ and display the results as a function of $L$. In Fig.~\ref{fig:1}(a) we see that the $q$-potential has the form of a tilted sawtooth function, with a positive local slope and negative discontinuities when a new mode is introduced. Thus, the system is extensive since, on average, the $q$-potential can be well approximated by a function of the form $\alpha L$.
An unexpected behavior shown in Fig.~\ref{fig:1}(a) is the decreasing average value of the entropy $S=q+M$ (leading to negative values of $S$) with increasing complexity. Although the entropy is increasing in the regions where it is continuous, the introduction of additional propagating modes leads to large negative jumps in the entropy. In Fig.~\ref{fig:1}(c)-(d), we see that both the optical temperature and the chemical potential are, for all practical purposes, almost constant and thus intensive.

In addition, in Fig.~\ref{fig:1}(b) we compute the optical pressure according to three different methods. The first is the direct evaluation of the exact expression of Eq.~(\ref{eq:PressMetal}). We observe that this leads to pressure singularities at the points where the new modes are introduced. These infinities are triggered from the $1/\varepsilon^{(M)}(L)$ terms that go to infinity as $L\rightarrow F(M)^+$. As we explained above, the presence of singularities is a shortcoming of the Rayleigh-Jeans distribution, which predicts positive power occupation numbers even at the cutoff between a propagating and an evanescent mode [$\varepsilon^{(M)}\rightarrow0^+$ in Eq.~(\ref{eq:epsilonmetallic})]. To resolve this problem, we can truncate the last term of Eq.~(\ref{eq:PressMetal}) (green dotted lines) that is responsible for the singularities. Alternatively, we can use the approximate expression of Eq.~(\ref{eq:papprox}) (red dashed-dotted lines). Both of these approximations are valid in the positive temperature regime.

%\section{Figure 2}
\begin{figure}
\centerline{
\includegraphics[width=\columnwidth]{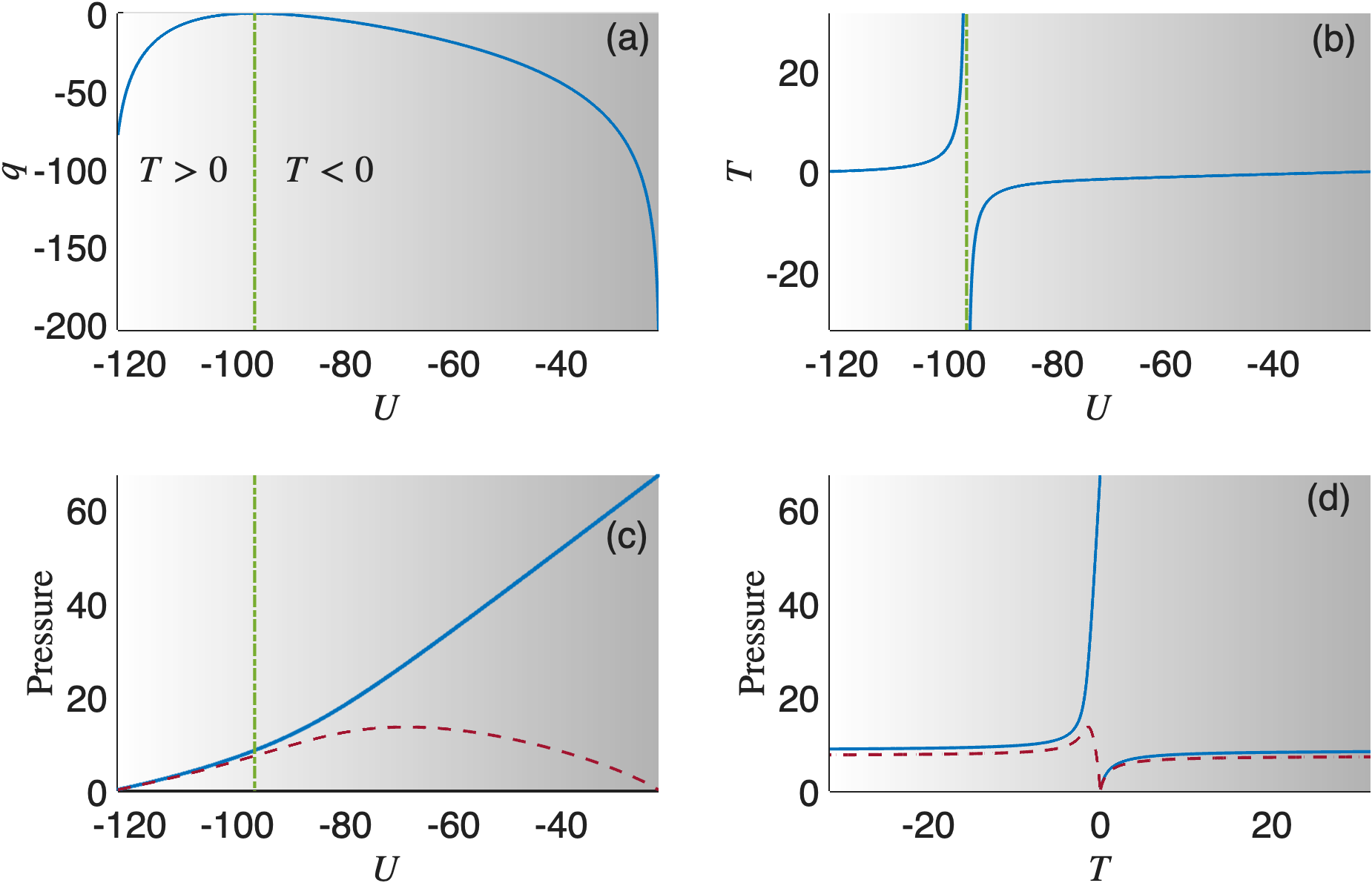}
}
\caption{Variation of (a) the $q$-potential, (b) the optical temperature as
functions of the internal energy, and the optical pressure as a function of (c) the internal energy and (d) the optical temperature for a metallic waveguide. The vertical green dashed-dotted lines separate the regions of positive and negative temperatures. In (c) and (d) the blue curve shows the exact calculation of the pressure using Eq.~(\ref{eq:PressMetal}) whereas the red dashed lines are computed using the asymptotic formula of Eq.~(\ref{eq:papprox}).}
\label{fig:2}
\end{figure}
In Fig.~\ref{fig:2} we present the thermal equilibrium properties of a metallic waveguide as a function of the internal energy. In particular, we keep constant the total power $N$ and the length of the waveguide $L$ and vary the internal energy to cover all acceptable values $\varepsilon^{(1)}N/\omega\le U\le \varepsilon^{(M)}N/\omega$. In Fig.~\ref{fig:2}(a) we depict the $q$-potential, which is equal to the entropy shifted downward by $M$. As expected at the maximum and minimum allowed values of $U$, the $q$ potential has a minimum with an infinite slope that indicates zero temperature [Fig.~\ref{fig:2}(b)]. On the other hand, when $q$ is maximum, the temperature becomes infinite. In Figs.~\ref{fig:2}(c)-(d), we present the optical pressure as a function of the internal energy and the optical temperature. Specifically, we use Eq.~(\ref{eq:PressMetal}) to compute the actual optical pressure, as well as the approximate expression given by Eq.~(\ref{eq:papprox}). In Fig.~\ref{fig:2}(d) we can see that the two formulas agree, except in the region where $T$ approaches the value $0^-$. However, by plotting the pressure as a function of the internal energy, we observe that Eq.~(\ref{eq:papprox}) can be considered as a good approximation of Eq.~(\ref{eq:PressMetal}) in a regime that covers about 30\% of the acceptable values of the energy, which is mainly associated with positive optical temperatures. As we already explained, once the distribution of the power occupation numbers becomes inverted (and the highest mode is associated with the highest energy), the asymptotic expression of Eq.~(\ref{eq:papprox}) starts to fail. In the limiting case of condensation to the highest energy state, the optical pressure of the approximate formula is minimized, in contrast to the exact formula that shows that the pressure takes its maximum value.

\begin{figure*}
\centerline{
\includegraphics[width=\textwidth]{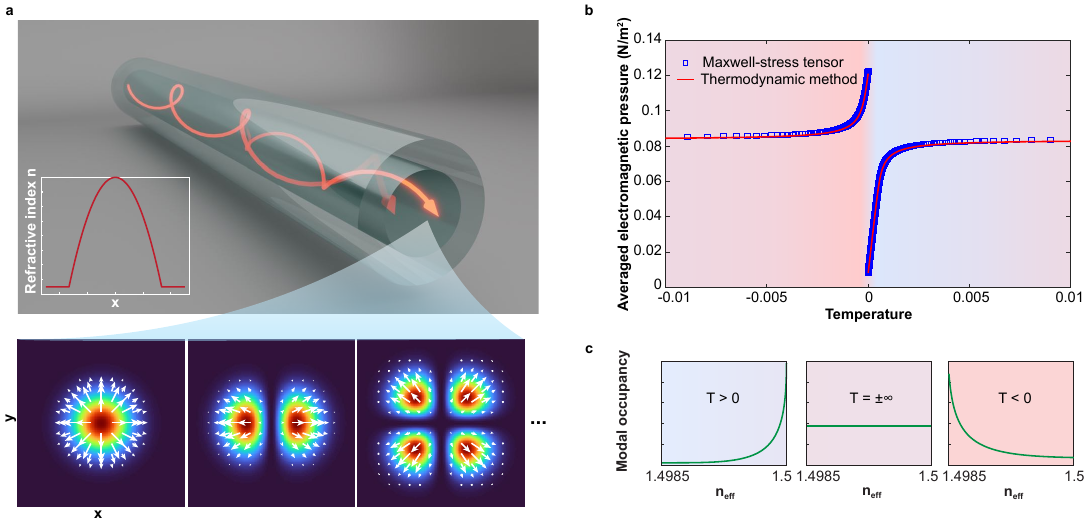}
}
\caption{(a) Light propagation in a weakly nonlinear multimode optical fiber with a graded-index profile. Different modes generate distinct radiation pressure distributions, as indicated by the white arrows. At equilibrium, the total pressure can be obtained by incoherently summing the contributions of all modes, weighted according to a Rayleigh-Jeans distribution. (b) Averaged electromagnetic pressure in the graded-index fiber shown in (a) under various initial excitation conditions. The red line indicates the pressure calculated using the optical thermodynamic method [Eq.~(\ref{eq:PressureGRIN2Dv2})], while the blue squares represent the pressure obtained from the Maxwell stress tensor. (c) Modal occupancies at thermal equilibrium across different temperature regimes.}
\label{fig:MaxwellGRIN}
\end{figure*}

\section{Thermodynamics in a graded-index fiber\label{sec:parabolic}}

Finally, we will examine the thermodynamics of a graded-index fiber associated with a parabolic index potential [Fig.~\ref{fig:MaxwellGRIN}(a)]. The general methodology developed here can be utilized to determine the pressure and the thermodynamic properties in other systems with smooth index variations as well as in Bose-Einstein condensates. The calculations and numerical results for the related problem of a one-dimensional waveguide with a parabolic index profile are presented in Appendix A. 

We assume paraxial wave propagation 
\begin{equation}
  i
  \frac{\partial\psi}{\partial z}
  +
  \frac1{2k}
  \left(
    \frac{\partial^2\psi}{\partial x^2}
    +
    \frac{\partial^2\psi}{\partial y^2}
  \right)
  +k_0n(r)\psi+\frac{\omega n_2}{c}|\psi|^2\psi
  = 0
  \label{eq:paraxial}
\end{equation}
in a medium with a finite-size parabolic-index potential
\begin{equation}
  n(r)
  =
  \begin{cases}
    -\Delta (r/R)^2,  & r\le R, \\
    -\Delta, & r> R,
  \end{cases}
\end{equation}
where $r=\sqrt{x^2+y^2}$ is the radial coordinate, $R$ is the radius of the waveguide, $\Delta$ is the refractive index contrast, $k_0=2\pi/\lambda$, $k=n_0k_0$, and $n_0$ is the refractive index of the cladding. In order to derive closed-form results, we assume that the modes and the respective propagation constants are identical to those of an untruncated parabolic potential and apply a modal cutoff. This is a good approximation for most of the modes besides the last few of them. Thus, it is expected to be highly accurate except in the case of high internal energies associated with negative temperatures, where higher-order modes are mostly occupied. Specifically, by defining the length scale
\[
  w_0 =
  \left(
    \frac{R}{k_0(2\Delta n_0)^{1/2}}
  \right)^{1/2}, 
\]
we can express the orthonormal Hermite-Gauss eigenmodes and the respective propagation constants as
\begin{equation}
  \psi^{(\ell_1,\ell_2)}=
  \frac{e^{-r^2/(2w_0^2)}}{ w_0(\pi2^{\ell_1}2^{\ell_2}\ell_1!\ell_2!)^{1/2}}
  H_{\ell_1}\left(\frac{x}{w_0}\right)
  H_{\ell_2}\left(\frac{y}{w_0}\right),
  \label{eq:HermiteGaussModes2}
\end{equation}
\begin{equation}
  \varepsilon^{(\ell_1,\ell_2)}
  =
  \frac{\ell_1+\ell_2+1}{k w_0^2},
  \label{eq:epsilonParabolicPot22}
\end{equation}
where $H_\ell$ are the Hermite polynomials.
The number of non-degenerate energy states $W$ is determined from the modal cutoff condition $\max(\varepsilon^{(\ell_1,\ell_2)}|_{\ell_1+\ell_2=W-1})\le k_0\Delta$ through
\[
  W 
  =
  \left\lfloor
    (k_0w_0)^2(n_0\Delta)
  \right\rfloor.
\]
Taking into account the modal degeneracies the total number of modes is given by $M=W(W+1)/2$.
Due to the continuous variation of the refractive index, the width of each mode is not globally characterized by the fiber radius $R$, but it varies along each direction with the modal indices $\ell_1,\ell_2$. However, due to the cylindrical symmetry, the root mean square radius of the beam at thermal equilibrium along both directions $\braket{x^2}^{1/2}$ and $\braket{y^2}^{1/2}$ is the same. Thus, we conclude that $\braket{r^2}=2\braket{x^2}=2\braket{y^2}$. For a supermode with indices $(\ell_1,\ell_2)$ we find that
\begin{equation}
  \langle r^2\rangle_{\ell_1,\ell_2}=
  \frac{
    \langle\psi^{(\ell_1,\ell_2)}|r^2|\psi^{(\ell_1,\ell_2)}\rangle
  }{
    \langle\psi^{(\ell_1,\ell_2)}|\psi^{(\ell_1,\ell_2)}\rangle
  }
  =
  w_0^2  (\ell_1+\ell_2+1).
  \label{eq:Var}
\end{equation}
We expand the optical wave into the supermodes [see Eq.~(\ref{eq:Psiexpansion})] with the coefficients $C^{(\ell_1,\ell_2)}$ satisfying the Rayleigh-Jeans distribution of Eq.~(\ref{eq:nl2}).
For Hermite-Gauss beams, the mean effective beam radius can be well approximated by $\tilde R^2=2\braket{r^2}$. Following the calculations, the mean effective area of the beam is given by
\begin{equation}
  \tilde A = \pi\tilde R^2 = 
  2\pi\frac{\braket{\Psi|r^2|\Psi}}{\braket{\Psi|\Psi}} =
  \frac{\omega U \pi R^2}{k_0\Delta N}=\sigma A,
\end{equation}
where
$
  \sigma
  =
  \omega U/(k_0\Delta N)
$
is the ratio of the effective beam area $\tilde A$ to the actual area of the waveguide $A$.

A question that naturally arises is whether the actual waveguide area $A$ or the effective beam area $\tilde A$ is the correct variable to describe the size of the system. First, let us point out that since $\sigma$ is an intensive parameter, the entropy is extensive with respect to both $(N,U,A)$ and $(N,U,\tilde A)$. Importantly,  both selections ($A$ or $\tilde A$) should provide the correct result for the average optical pressure.
Specifically using the pressure definition $p_{A}=T(\partial q /\partial A)_{\alpha,\beta}$ and changing variables from $A$ to $\tilde A$, we derive the following expression for the ratio of the average thermodynamic pressures
\[
    p_A/p_{\tilde A} = \tilde A/A.
  \]
Taking into account that $\tilde A$ should describe the effective area of the beam where almost all the power resides, cropping the remaining part should result in a pressure ratio that is inversely proportional to their area. We conclude that both $p_A$ and $p_{\tilde A}$ correctly describe the mean optical pressure as long as $\tilde A$ contains most of the beam power. 

The limiting values of $\tilde A$ are obtained from the condensation limits where all the power is concentrated either in the lowest $\ell_1=\ell_2=0$ or occupies the highest modes $\ell_1+\ell_2=W-1$ with the same, on average, amount of power and thus
\[
  \pi w_0^2
  \le
  \tilde A
  \le
  \pi Ww_0^2.
\]
We can now follow the thermodynamic calculations in a similar fashion as in Section~\ref{sec:metal} with a partition function $q=q(\alpha,\beta,\tilde A)$. The optical pressure is the conjugate variable to $\tilde A$, leading to
\begin{equation}
  p_{\tilde A} = T
  \left(
    \frac{\partial q}{\partial \tilde A}
  \right)_{\alpha,\beta}
  =
  -\sum_{\ell=0}^{M-1}
  \frac{n^{(\ell)}}{\omega}
  \frac{\dif\varepsilon^{(\ell)}}{\dif\tilde A}
  =
  -\sum_{\ell=0}^{M-1}
  \frac{n^{(\ell)}}{\omega\sigma}
  \frac{\dif\varepsilon^{(\ell)}}{\dif A}.
  \label{eq:PressureGRIN2Dv1}
\end{equation}
Note that in this case we have two transverse directions, and thus the force is measured in units of force per unit area.
Substituting into Eq.~(\ref{eq:PressureGRIN2Dv1}) the propagation constants given by Eq.~(\ref{eq:epsilonParabolicPot22}), we obtain the following very simple exact expression for the optical pressure
\begin{equation}
  p_{\tilde A} = \frac{U}{2\tilde A},
  \label{eq:PressureGRIN2Dv2}
\end{equation}
which is proportional to the internal energy and inversely proportional to the effective beam area. In a similar fashion to the previous section about metallic waveguides, we observe discontinuities in the thermodynamic parameters ($q,p,T,\mu$) when we proportionally increase $(N,U,A)$. The discontinuities take place when a new mode is added to the system. However, in contrast to the results of Section~\ref{sec:metal}, due to the functional form of the propagation constants, here we do not observe pressure singularities. Specific details can be found in Appendix~\ref{sec:parabolic1d}, where we examine the corresponding one-dimensional problem.

It is worth noting that Eq.~(\ref{eq:PressureGRIN2Dv2}) represents an important result: it enables the electrodynamic pressure to be determined effortlessly once the system reaches thermal equilibrium. Owing to the unique characteristics of the propagation constants in graded-index fibers, the electromagnetic pressure depends solely on the internal energy, as shown in Eq.~(\ref{eq:PressureGRIN2Dv2}), closely resembling the behavior observed in an ideal gas~\cite{pathr-2011}. To illustrate the theoretical results, we present numerical simulations for a weakly guided graded-index fiber with a radius of $a = 25$ \unit{\micro\meter}, as schematically depicted in Fig.~\ref{fig:MaxwellGRIN}(a). The core and cladding refractive indices are $n_1=1.46$ and $n_2=1.4454$, respectively. The fiber operates at a wavelength of $\lambda=1064$ \unit{\nano\meter} and supports $M=120$ Hermite-Gaussian modes~\cite{okamo-2006}. These modes produce distinct optical radiation pressure distributions, as shown in Fig.~\ref{fig:MaxwellGRIN}(a). Our simulation results, shown in Fig.~\ref{fig:MaxwellGRIN}(b), demonstrate excellent agreement between the two methods across the entire temperature range. In Fig.~\ref{fig:MaxwellGRIN}(c), we plot the corresponding Rayleigh-Jeans distributions under positive and negative temperature conditions, where lower- and higher-order modes are preferentially populated, respectively. Meanwhile, at infinite temperatures ($T\rightarrow\pm\infty$), power equipartition ensues among modes. Notably, as $T\rightarrow 0^-$, the electrodynamic pressure increases due to the increased population of higher-order modes, while as $T\rightarrow 0^+$, lower-order modes are favored, resulting in a decrease in electrodynamic pressure. This behavior is consistent with the observations in Fig.~\ref{fig:MaxwellGRIN}(b).

Optical pressure in graded-index multimode fibers can induce optomechanical deformations and, through the photoelastic effect, measurable changes in the refractive-index distribution. Here, we provide an introductory discussion based on order-of-magnitude estimates, leaving a comprehensive treatment to future work. The calculations are presented in Appendix~\ref{sec:photoelastic}.
  Using parameters relevant to silica graded-index fibers ($R=26.05$ \unit{\mu\meter}, $n_0=1.46$, $\Delta=1.44\times10^{-2}$, $\lambda=1064$ \unit{\nano\meter}, $E=73.1$ \unit{\giga\pascal}, $\nu=0.17$, $p_e=0.22$), we find that the photoelastically induced, mode-dependent shifts of the propagation constants per unit power  are of the order of $10^{-7}$ \unit{\per\watt\per\meter}. For example, substituting the above parameters in Eq.~(\ref{eq:deltabeta}) of Appendix~\ref{sec:photoelastic} we obtain $|\delta\beta^{(0,0)}-\delta\beta^{(1,1)}|\simeq1.09\times10^{-7}P$ \unit{\per\meter}. For the representative input power $P=100$ \unit{\kilo\watt} and  propagation distance $L=100$ \unit{\meter} the accumulated differential phase shift is of the order of $1$ \unit{\radian}.

\section{System extensivity\label{sec:extensivity}}

In classical thermodynamics, we identify the extensive natural variables of entropy to be the energy $\hat U$, the number of particles $\hat N$, and the volume $\hat V$. This means that if a system has entropy $S(\hat U,\hat N,\hat V)$ and we proportionally scale all the extensive natural variables by $\lambda$ while keeping all the intensive $\hat\xi$ fixed, the entropy will also proportionally increase to $S(\lambda\hat U,\lambda\hat N,\lambda\hat V;\hat \xi) = \lambda S(\hat U,\hat N,\hat V;\hat\xi)$. In optical thermodynamics different quantities may, at first glance, appear to be candidates for the third extensive natural variable of entropy. However, the final selection has to rely on the fact that the entropy is extensive when we scale all the extensive natural variables of entropy while keeping all the remaining parameters constant [see Eq.~(\ref{eq:entropy_ext})]. 

Let us summarize our findings about system extensivity. The results are also illustrated schematically in Fig.~\ref{fig:settings}. The two extensive natural variables of entropy in multimode optical systems are always the two conservation laws; namely the total power $N$ and the internal energy (or energy per unit of propagation distance, or momentum flow) $U$. The third natural variable of entropy depends on the system under investigation. Specifically, we have found that it is easily recognizable as a parameter ($\xi_1$) of the propagation constants, by requiring system extensivity through Eq.~(\ref{eq:entropy_ext}): $S(\lambda N,\lambda U,\lambda\xi_1;\xi_2,\ldots,\xi_J)=\lambda S(N,U,\xi_1;\xi_2,\ldots,\xi_J)$.
For example, in discrete settings, such as those studied in Sections~\ref{sec:array}-\ref{sec:SSH}, the third extensive natural variable of entropy is the number of primitive cells $\tilde M$. On the other hand, in continuous systems, including those examined in Section~\ref{sec:metal}-\ref{sec:parabolic}, the last extensive natural variable of entropy is the size of the system $\Omega$ (length, area, or volume, depending on the system dimensionality), or even the effective beam size $\tilde\Omega$ (if the system does not have abrupt boundaries).

In discrete systems the extensive natural variables of the entropy are $(U,N,\tilde M)$.
Increasing $\tilde M$ corresponds to replicating identical unit cells at fixed intensive parameters. The rest of the parameters, such as the spacing between waveguides, the lattice arrangement and geometry, the coupling coefficients, and even the boundary conditions, should remain invariant/intensive as we proportionally increase $(U,N,\tilde M)$. The total number of modes is not an appropriate extensive variable, because changing $M$ by one can modify the spectral/topological properties of the system (for example the topological edge states in an SSH lattice).
The question that naturally arises is why we selected $\tilde M$ over a parameter that describes the size of the system ($L$ or more generally $\Omega$). First, let us point out that we did not select $\tilde M$ but it was naturally provided from the propagation constants. More importantly, we see that  increasing $\tilde M$ leads to the proportional increase of the size of the system $L$. However, the opposite is not always true: Increasing the size of the system $L$ does not necessarily imply the proportional increase of the number of primitive cells $\tilde M$. Since $L(\tilde M,s) = s\tilde M$ we can increase $L$ either by changing $\tilde M$ or by changing the spacing between primitive cells $s$. Thus, the entropy should satisfy $S(\lambda U,\lambda N,\lambda\tilde M;\xi)=\lambda S(U,N,\tilde M;\xi)$ where $\xi$ are intensive parameters. For example in general SSH lattices $\xi=\{s_1,s_2,\beta_1,\beta_2\}$, where $\beta_1,\beta_2$ are the propagation constants of the two sublattices.

Similar arguments hold for bulk systems. Here the third extensive natural variable of the entropy is the size of the system $\Omega$ (length $L$, area $A$, or volume $V$) or the effective size $\tilde\Omega$, meaning that the entropy is homogeneous of degree 1 with respect to $(U,N,\Omega)$ while keeping the rest of the parameters constant. We ask again the same question: Why did we select $\Omega$ over $M$? Again, we emphasize that $\Omega$ was not selected but emerges naturally from the propagation constants. What is more crucial, is that increasing $\Omega$ (or $\tilde\Omega$) proportionally increases $M$ (Weyl's law for elliptic differential operators). However, increasing the number of modes $M$ does not necessarily proportionally increase the size of the system $\Omega$: The number of modes does not depend only on the size of the system but also on other parameters such as the refractive index distribution. For example, for step-index waveguides $S(\lambda U,\lambda N,\lambda A;n_1,n_2) = \lambda S(U,N,A;n_1,n_2)$, where $n_1$, $n_2$ are the refractive indices of the core and the cladding.

\section{Optical pressure\label{sec:pressure}}

We find that the following expression
\begin{equation}
  p =
  - \sum_{\ell=1}^M
  \frac{n^{(\ell)}}{M\omega}
  \left(
    \frac{\partial\varepsilon^{(\ell)}}{\partial s}
  \right)_{\alpha,\beta,M,\overline\xi_j}
  = 
  - \sum_{\ell=1}^M
  \frac{n^{(\ell)}}{\omega}
  \left(
    \frac{\partial\varepsilon^{(\ell)}}{\partial L}
  \right)_{\alpha,\beta,M,\overline\xi_j}
  \label{eq:FinalPressure}
\end{equation}
for the pressure has a global character and thus can be applied to derive the optical forces in both discrete and continuous optical settings. This formula was originally derived in Ref.~\cite{efrem-cp2022}. Note however, that according to the system under investigation the parameters and interpretation of Eq.~(\ref{eq:FinalPressure}) have to be accordingly tuned.
The middle part above applies to discrete systems, whereas the last expression should be utilized in continuous settings. The main expressions related to optical pressure are summarized in Fig.~\ref{fig:PressureResults}.

In particular, in monoatomic photonic lattices (Section~\ref{sec:array}) the second part can be directly applied with $s$ being the spacing between adjacent waveguides and $M$ being the number of waveguides. In polyatomic lattices, $s$ is the length of the primitive cell and we have to replace $M$ with the number of primitive cells $\tilde M$. In this case, Eq.~(\ref{eq:FinalPressure}) represents the average optical pressure for all the bonds along the array. In addition, in SSH diatomic lattices examined in Section~\ref{sec:SSH}, it can also be used to derive the intercell and intracell pressures [Eq.~(\ref{eq:pjSSH})]. Here, $s$ is replaced with $s_j$, $j=1,2$ and $\overline \xi_j=s_{3-j}$. Finally, by replacing $s$ with $r=s_1-s_2$ and keeping constant $\overline\xi_j=s$, we can compute the pressure differential term from Eq.~(\ref{eq:strain}). Such generalizations can also be utilized in other types of polyatomic lattices. 

In bulk systems, following the discussion of Section~\ref{sec:extensivity}, the size of the system $L$ is the third extensive natural variable of entropy.
The number of modes $M(L)$ and thus $s=L/M(L)$ are uniquely determined from the size of the system $L$ (keeping the rest of the parameters constant). Thus, the middle expression for the pressure in Eq.~(\ref{eq:FinalPressure}) does not apply and we have to utilize the last term that involves total length of the system $L$. This formula was used for metallic waveguides in Section~\ref{sec:metal}.

In systems with refractive-index variations that are spatially inhomogeneous, such as the graded-index configurations studied in Section~\ref{sec:parabolic} and Appendix~\ref{sec:parabolic1d}, the optical forces are not only exerted close to the geometric interfaces, but are distributed along the beam footprint on the areas with index gradients.  This is apparent from the following expression for the time-averaged electromagnetic force density $\langle\bm{f}(\bm r)\rangle\approx -(\epsilon_0/2)n(\bm r)|\bm E(\bm r)|^2\nabla n(\bm r)$ (up to electrostrictive terms). In systems with spatially varying index, the beam size $\tilde L$ (or more generally $\tilde\Omega$) can be smaller than the actual size of the fiber or waveguide $L$ (or $\Omega$). In our calculations we compute the beam size from the variance $\braket{x^2}$ although different selections are also possible. Outside this effective beam area $\tilde\Omega$ the optical forces are almost zero.  Substituting into Eq.~(\ref{eq:FinalPressure}) $L\rightarrow\tilde L$, or  $A\rightarrow\tilde A$ (2D) we compute larger mean pressure since we divide the same total force by a smaller area. Both expressions for the pressure $p_{\tilde A}$ and $p_A$ are correct, however $p_{\tilde A}$ avoids averaging the total force over regions outside the effective beam footprint.

Depending on the system under study, different exact or asymptotic expressions for the optical pressure are derived. For example, in discrete monoatomic systems an exact expression for the pressure is given by Eq.~(\ref{eq:pArraySimplified}) [Ref.~\cite{efrem-cp2022}], whereas in graded-index systems we have derived Eq.~(\ref{eq:PressureGRIN2Dv2}). Both of these expressions are proportional to the internal energy and inversely proportional to the number of modes [Eq.~(\ref{eq:pArraySimplified})] or the effective size of the system [Eq.~(\ref{eq:PressureGRIN2Dv2})]. In metallic waveguides Eq.~(\ref{eq:papprox}), which was originally derived in Ref.~\cite{ren-prl2023}, is very accurate for positive temperatures.

\section{Limitations of the Rayleigh-Jeans distribution\label{sec:limitations}}

We have identified deviations in the result obtained using the Rayleigh-Jeans distribution as compared to the expected behavior. They originate from the fact that optical thermodynamics describes systems with a finite number of modes. The first one is the presence of positive power occupation numbers when a mode is at its cutoff. However, such a mode has zero propagation constant and thus cannot carry power. Assume that a parameter $\xi$ that controls, for example, the refractive index contrast or the size of a system, can be tuned. Specifically, when $\xi<\xi_0$ the system supports $M-1$ modes, whereas for $\xi\ge\xi_0$ the system has $M$ modes. Exactly at the cutoff $\xi=\xi_0$, due to continuity, we expect that the power occupation number $n^{(M)}$ should be identical to zero. However, due to the functional form of Eq.~(\ref{eq:nl2}) the power occupation numbers are always positive even at the cutoff. 

The presence of a positive power occupation number at the cutoff results in discontinuities in most of the resulting thermodynamic quantities. Specifically, as we can see in Figs.~\ref{fig:1} and~\ref{fig:3}, the temperature, the chemical potential and the $q$-potential are discontinuous functions. Note that from Eq.~(\ref{eq:dqMMetal}) we see that if $n^{(\ell)}$ vary continuously with $L$, then the $q$-potential will be a continuous function. 

More importantly, we expect that the entropy $S=q+M$ will increase proportionally with the extensive parameters of the system. Although the entropy is linearly increasing in the regions where it is continuous, there are cases [for example in metallic waveguides -- see Fig.~\ref{fig:1}(a)] where the negative discontinuities at the locations where additional modes are introduced lead to an, on average, decreasing entropy. These jumps do not affect the on average homogeneity of degree one of the entropy, however we obtain an unphysically negative and decreasing entropy with increasing complexity of the problem.

A discrepancy that affects physically measurable quantities is already presented in Fig.~\ref{fig:1}(b). In the case of metallic waveguides, at the locations where additional modes are introduced ($L=F(m)=m\pi/k$) the exact expression for the pressure goes to infinity. This problem only affects the values of the pressure at the right limit of the discontinuities $L=\lim_{\epsilon\rightarrow0^+}(F(m)+\epsilon)$. As we can see from Eq.~(\ref{eq:PressMetal}), this problem can be resolved if $n^{(m)}$ satisfies a distribution that goes to zero faster than $\varepsilon^{(m)}$ when the $m$th mode is introduced to the system.

However, other than the above limitations, the Rayleigh-Jeans distribution in optical thermodynamics is an excellent model to describe the thermal equilibrium conditions of weakly nonlinear multimode optical systems. The power occupation numbers show very good agreement with numerical and experimental results. The resulting temperature and chemical potential are uniquely obtained from the energy and the power conservation laws. Note that we expect that the above discrepancies cannot be resolved by utilizing a Bose-Einstein distribution. The main reason is that the Bose-Einstein statistics still predict non-zero power occupation numbers at the cutoff.

\section{Conclusions}

In conclusion, we have shown that in optical thermodynamics each multimode nonlinear optical platform is characterized by a distinct set of extensive and intensive parameters. Thus, the final thermodynamic description can differ substantially from one system to another. We have provided a unified methodology to identify the relevant thermodynamic parameters and derive the thermodynamic relations and the optical pressure, which is directly applicable to any type of multimode weakly nonlinear optical setting. We apply our results to a variety of optical systems, including lattices with a different number of ``atoms'' per unit cell, waveguides, as well as systems with smooth index variations such as graded-index fibers. We expect that our work is important in analyzing different optical settings, as well as understanding and explaining a number of fundamental aspects that lie in the core of optical thermodynamics.

%%%%

\begin{acknowledgments}

This work was supported by the Air Force Office of Scientific Research (AFOSR) Multidisciplinary University Research Initiative (MURI) (award no. FA9550-20-1-0322), AFOSR MURI (award no. FA9550-21-1-0202), ONR MURI (award no. N00014-20-1-2789), the Army Research Office (W911NF-23-1-0312), the Department of Energy (DE-SC0022282), the Department of Energy (DE-SC0025224), MPS Simons collaboration (Simons grant no. 733682), US Air Force Research Laboratory (FA86511820019) and AFRL - Applied Research Solutions (S03015) (FA8650-19-C-1692).

\end{acknowledgments}

\section*{Data Availability}

The data supporting this study's findings are available within the article.

\appendix
\section{Thermodynamics in a one-dimensional graded-index waveguide \label{sec:parabolic1d}}

We investigate the thermodynamics of a one-dimensional waveguide in the paraxial regime
\begin{equation}
  i
  \frac{\partial\psi}{\partial z}
  +
  \frac1{2k}
  \frac{\partial^2\psi}{\partial x^2}
  +k_0n(x)\psi+\frac{\omega n_2}{c}|\psi|^2\psi
  = 0
  \label{eq:app:paraxial}
\end{equation}
in a medium with a finite size parabolic index potential
\begin{equation}
  n(x)
  =
  \begin{cases}
    -\Delta (2x/L)^2,  & |x|\le L/2, \\
    -\Delta, & |x|> L/2,
  \end{cases}
\end{equation}
where $L$ is the diameter of the waveguide and $\Delta$ is the refractive index contrast. In Eq.~(\ref{eq:app:paraxial}) we consider Kerr nonlinearity, $k_0=2\pi/\lambda$, $k=n_0k_0$, and $n_0$ is the refractive index of the cladding. In order to derive closed-form results, we assume that the modes and the respective propagation constants are identical to those of an untruncated parabolic potential and apply a modal cutoff. This is a good approximation for most of the modes besides the last few of them. Thus, it is expected to be highly accurate except in the case of high internal energies associated with negative temperatures, where higher-order modes are mostly occupied. Specifically, by defining the width 
\[
  w_0 =
  \left(
    \frac{L}{2k_0(2\Delta n_0)^{1/2}}
  \right)^{1/2}, 
\]
we can express the orthonormal Hermite-Gauss eigenmodes and the respective propagation constants as
\begin{equation}
  \psi^{(\ell)}(x)=
  \frac{e^{-x^2/(2w_0^2)}}{(\pi^{1/2}2^\ell\ell!w_0)^{1/2}}
  H_\ell\left(\frac{x}{w_0}\right),
  \label{eq:app:HermiteGaussModes2}
\end{equation}
\begin{equation}
  \varepsilon^{(\ell)}
  =
  \frac{1}{k w_0^2}
  \left(
    \ell+\frac12
  \right),
  \label{eq:app:epsilonParabolicPot22}
\end{equation}
where $H_\ell$ are the Hermite polynomials and $\ell=0,1,\ldots,M-1$. 
The modal cutoff value $M$ is determined from the condition $\max(\varepsilon^{(M-1)})\le k_0\Delta$ leading to the following expression for the number of modes
\[
  M
  =
  \left\lfloor
    (k_0w_0)^2\Delta n_0+\frac12
  \right\rfloor.
\]
Due to the continuous variation of the refractive index, the width of each mode is not globally characterized by the waveguide diameter $L$, but it varies with the modal index $\ell$. Specifically, as the mode number increases, the spatial extent of the mode also increases. The effective size of each mode can be estimated from the variance of $x$. Since the eigenmodes are either even or odd, we have that $\langle x\rangle_\ell=0$. Thus the variance is equal to the second-order moment and is given by
\begin{equation}
  \operatorname{Var}_\ell(x)
  =
  \langle x^2\rangle_\ell=
  \frac{
    \langle\psi^{(\ell)}|x^2|\psi^{(\ell)}\rangle
  }{
    \langle\psi^{(\ell)}|\psi^{(\ell)}\rangle
  }
  =
  \left(
    \ell+\frac12
  \right)
  w_0^2.
  \label{eq:app:Var}
\end{equation}
From Eq.~(\ref{eq:app:Var}) we see that the rms size of each mode scales as $\ell^{1/2}$. The beam variance in thermal equilibrium can be computed by using a modal expansion $\ket{\Psi}=\sum_{\ell=0}^{M-1}C^{(\ell)}\ket{\ell}$ with the coefficients $C^{(\ell)}$ satisfying the Rayleigh-Jeans distribution $n^{(\ell)}=T/(\varepsilon^{(\ell)}/\omega-\mu)$. For Hermite-Gauss beams, we find that the beam diameter can be well approximated by $\tilde L=(8\langle x^2\rangle)^{1/2}$. Following the calculations we derive the following relation
\begin{equation}
  \tilde L^2 = 
  8\frac{\braket{\Psi|x^2|\Psi}}{\braket{\Psi|\Psi}} =
  \frac{\omega UL^2}{k_0\Delta N}=\sigma^2L^2,
\end{equation}
where
\begin{equation}
  \sigma
  =
  \frac{\tilde L}{L}
  =
  \sqrt{\frac{\omega U}{k_0\Delta N}}.
  \label{eq:app:sigma}
\end{equation}
is the ratio of the effective beam diameter $\tilde L$ to the actual diameter of the waveguide $L$.

\begin{figure}
\centerline{
\includegraphics[width=\columnwidth]{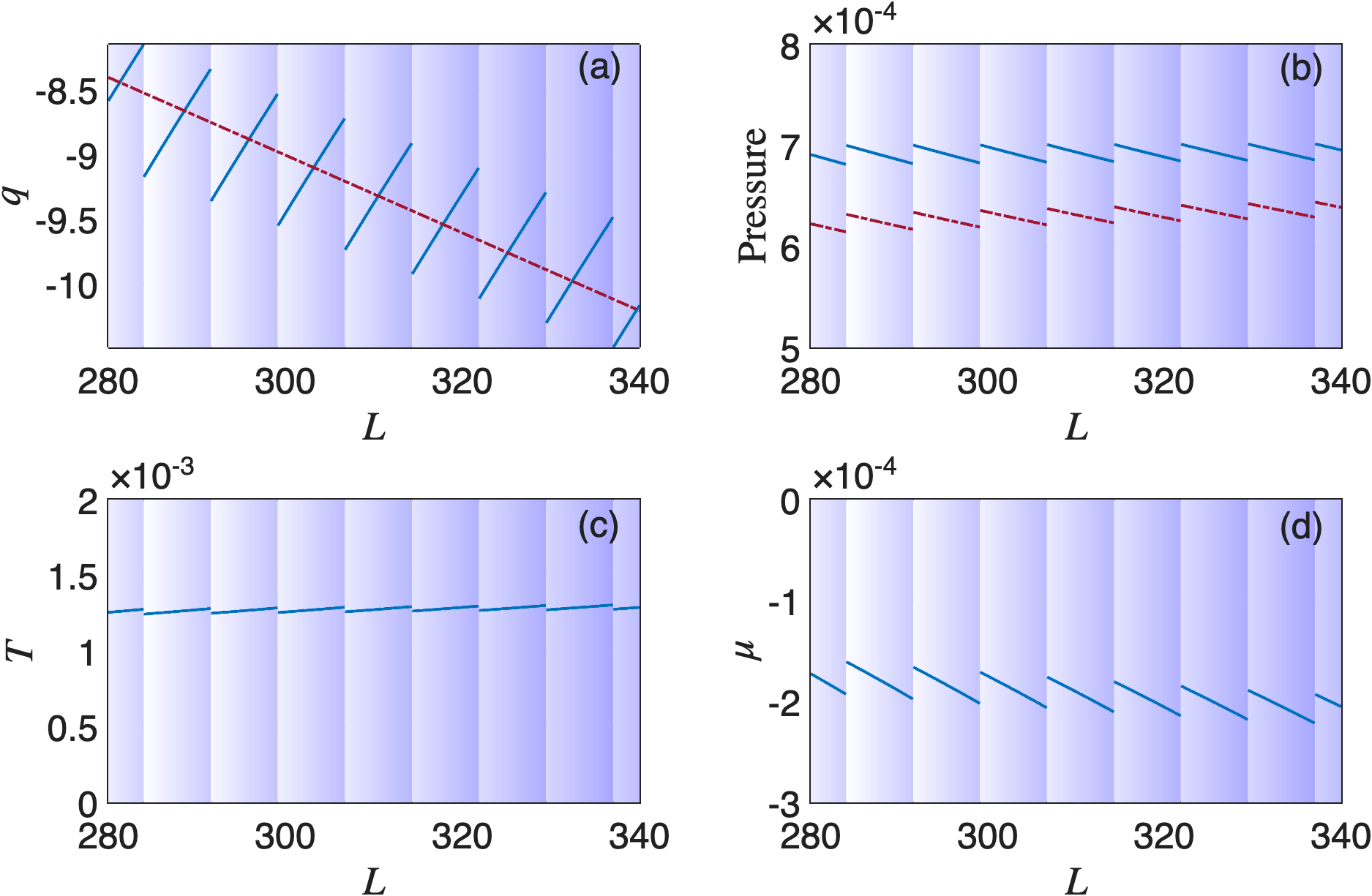}
}
\caption{Proportional variation of the set of extensive natural variables of entropy ($L, N, U$) for a one-dimensional parabolic potential. To display the effect of the introduction of new modes, in the horizontal axis we vary $L$ for a narrow range of values. (a) The $q$-potential shown with blue lines is extensive and thus, on average, fits well with the red dashed dotted line that passes through the origin. We see that $q$ is piecewise continuous with discontinuities at the locations where new modes are introduced. (b) Exact optical pressure using Eq.~(\ref{eq:PressureGRIN}) is presented with blue lines, whereas the dashed-dotted red line depicts the approximate calculation for the pressure using Eq.~(\ref{eq:papprox}). In (c) and (d), we observe that the temperature and the chemical potential remain almost invariant as $L$ varies, verifying their intensive character.}
\label{fig:3}
\end{figure}

As in the two-dimensional case of graded-index fibers studied in Section~\ref{sec:parabolic}, we see that entropy is extensive with respect to both $(N,U,L)$ and $(N,U,\tilde L)$. Specifically using the definition of the pressure $p_L=T(\partial q /\partial L)_{\alpha,\beta}$, we derive the following expression for the ratio of the average thermodynamic pressures
\[
  p_L/p_{\tilde L} = \tilde L/L. 
\]
Taking into account that $\tilde L$ should describe the effective diameter of the beam where almost all the power resides, cropping the remaining part should result to a pressure ratio that is inversely proportional to their length. We conclude that both $p_L$ and $p_{\tilde L}$ correctly describe the mean optical pressure as long as $\tilde L$ contains most of the beam power. 

The limiting values of $\tilde L$ are obtained from the condensation limits where all the power is concentrated either in the lowest $\ell=0$ or in the highest mode $\ell=M-1$ and thus
\[
  4w_0^2
  \le
  \tilde L^2
  \le
  8
  \left(M-\frac12\right) 
  w_0^2.
\]
We can now follow the thermodynamic calculations in a similar fashion as in Section~\ref{sec:general} with a partition function $q=q(\alpha,\beta,\tilde L)$. Thus, Eq.~(\ref{eq:PressureGRIN2Dv1}) for the pressure becomes
\begin{equation}
  p_{\tilde L} = T
  \left(
    \frac{\partial q}{\partial \tilde L}
  \right)_{\alpha,\beta}
  =
  -\sum_{\ell=0}^{M-1}
  \frac{n^{(\ell)}}{\omega}
  \frac{\dif\varepsilon^{(\ell)}}{\dif\tilde L}
  =
  -\sum_{\ell=0}^{M-1}
  \frac{n^{(\ell)}}{\omega\sigma}
  \frac{\dif\varepsilon^{(\ell)}}{\dif L}.
  \label{eq:pressureGRIN}
\end{equation}
Substituting into Eq.~(\ref{eq:pressureGRIN}) the propagation constants given by Eq.~(\ref{eq:app:epsilonParabolicPot22}), we obtain the following very simple exact expression for the optical pressure
\begin{equation}
  p_{\tilde L} = \frac{U}{\tilde L},
  \label{eq:PressureGRIN}
\end{equation}
which is proportional to the internal energy and inversely proportional to the effective beam size. The rest of the results are similar to those obtained in Section~\ref{sec:metal}, with the substitution $L\rightarrow\tilde L$ and do not need to be repeated.

% \section{Figure 3}

In Fig.~\ref{fig:3}, we depict $q$, $p$, $T$, and $\mu$ as a function of $L$, when the extensive natural variables of entropy ($U,N,\tilde L$) are proportionally increased. We see that the $q$-potential is extensive [Fig.~\ref{fig:3}(a)] and thus, on average, it can be approximated by a straight line passing through the origin. In Fig.~\ref{fig:3}(b) the pressure is computed using both the exact formula given by Eq.~(\ref{eq:PressureGRIN}) as well as the approximate expression given by Eq.~(\ref{eq:papprox}). 
For the selected parameters the optical temperature is positive, and both formulas give comparable results, with the asymptotic formula slightly underestimating the resulting optical pressure. In contrast to the case of metallic waveguides, here we do not observe any singularities in the computation of the pressure. This is an outcome of the functional form of the propagation constants. Finally, in Figs.~\ref{fig:3}(c)-(d) we plot the optical temperature and the chemical potential both of which, due to their intensive character, remain almost constant as expected.

\begin{figure}
\centerline{
\includegraphics[width=\columnwidth]{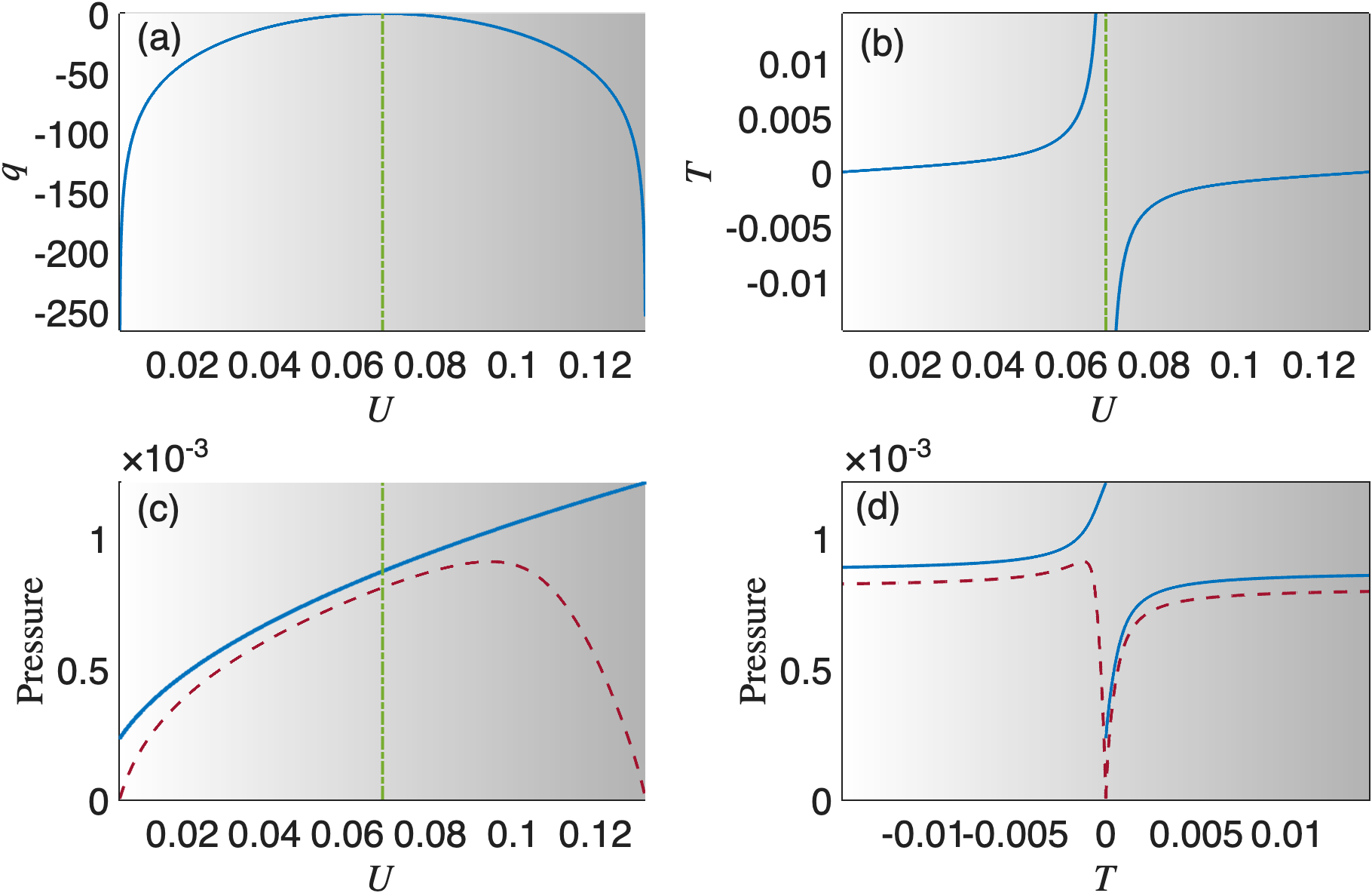}
}
\caption{Variation of (a) the $q$-potential, (b) the optical temperature, and (c) the optical pressure as a function of the internal energy in a one-dimensional graded-index arrangement. (d) Pressure as a function of the optical temperature. The power and the size of the waveguide are kept constant. The vertical green dashed-dotted lines separate the regions of positive and negative temperatures. In (c) and (d) the blue curve shows the exact calculation of the pressure using Eq.~(\ref{eq:PressureGRIN}) whereas the red dashed lines are computed using Eq.~(\ref{eq:papprox}).}
\label{fig:4}
\end{figure}

In Fig.~\ref{fig:4} we vary only the internal energy, while keeping constant values for the power and the size of the fiber $L$. Specifically, the internal energy takes values in the maximum allowed range $\varepsilon^{(0)}N/\omega\le U\le \varepsilon^{(M-1)}N/\omega$. As we can see in Fig.~\ref{fig:4}(a), the $q$-potential is minimum with an infinite slope at the condensation limits and reaches a maximum in between. The slope of the $q$-potential determines the temperature which is shown in Fig.~\ref{fig:4}(b). Two methods for the computation of the pressure are compared in Fig.~\ref{fig:4}(c). We observe that the asymptotic formula of Eq.~(\ref{eq:papprox})
fails to approximate Eq.~(\ref{eq:PressureGRIN}) at relatively large values of the internal energy associated with negative temperature. Finally, in Fig.~\ref{fig:4}(d) we present the optical pressure as a function of the temperature.

\section{Photoelastic calculations\label{sec:photoelastic}}

There are a few key ingredients needed to relate optical pressure to measurable changes in the radius or the index distribution of the fiber. Since the contribution associated with strain-induced changes of the fiber radius is expected to be smaller, we focus here on the photoelastic modification of the refractive index. In 3D isotropic linear elasticity, the stress $\sigma$ and strain $\tilde\varepsilon$ tensors are related through
  \begin{equation}
    \tilde\varepsilon_{ij}=\frac{1+\nu}{E}\sigma_{ij}-\frac\nu E\tr(\sigma)\delta_{ij},
    \label{eq:stress-strain}
  \end{equation}
  where $E$ is Young's modulus, and $\nu$ is Poisson's ratio. We denote the strain tensor by  $\tilde\varepsilon$ in order to avoid confusion with the propagation constants. The static force balance between stress and external force density requires
  \begin{equation}
    \nabla\cdot\sigma+\bm f=0.
    \label{eq:stress-force}
  \end{equation}
  Optical force density can in principle be computed from the Maxwell stress tensor. For analytical estimates, however, it is useful to employ the local approximation
  \begin{equation}
    \braket{\bm f(\bm r)}
    \approx
    -\frac{\epsilon_0}{2}
    n(\bm r) |\bm E(\bm r)|^2\nabla n(\bm r)
    \label{eq:optical_forces}
  \end{equation}
  which applies to a linear, isotropic dielectric with slowly varying material parameters. This expression shows that optical forces arise from spatial gradients of the refractive index and are therefore naturally relevant in graded-index fibers. 
  Finally, the strain field modifies the refractive index through the photoelastic effect. In a scalar isotropic approximation
  \begin{equation}
    \delta n(\bm r) =
    -
    \frac{n_0^3}{2}p_e\tr\tilde\varepsilon(r)
    \label{eq:index_contrast}
  \end{equation}
  where $p_e$ is an effective photoelastic constant.

  To get an order of magnitude estimate of the mode-dependent changes in the propagation constants,  $|\delta\beta^{(\ell_1,\ell_2)}-\delta\beta^{(\ell_1',\ell_2')}|$ we approximate the intensity profile by a Gaussian beam with radius $\tilde R$
  \begin{equation}
    I(r) = \frac{P}{\pi\tilde R^2}\exp\left(
      -\frac{r^2}{\tilde R^2}
    \right). 
  \end{equation}
  From Eq.~(\ref{eq:optical_forces}) and matching the work performed by the effective mean pressure $p_{\tilde A}$ to that associated with the radial force density $f_r(r)$ over the beam cross-section, we obtain 
  \[
    f_r(r) =
    \frac{2p_{\tilde A}}{\tilde R^2}
    r
    \exp\left(
      -\frac{r^2}{\tilde R^2}
    \right).
  \]
  Then from Eq.~(\ref{eq:stress-force}) we can compute the stress, and subsequently, through Eq.~(\ref{eq:stress-strain}) the strain under plane-stress assumption $\sigma_{zz}=0$. Specifically, for axisymmetric stress, the differential equation for  $g(r) = \tilde\varepsilon_{rr}+\tilde\varepsilon_{\theta\theta}$  satisfies
  \[
    g'(r) = -\frac{1-\nu^2}{E}f_r(r)
  \]
  which can be analytically integrated
  \[
    g(r) = \frac{1-\nu^2}{E} p_{\tilde A}
    \exp\left(
      -\frac{r^2}{\tilde R^2}
    \right)
  \]
  Thus, the trace of the strain is given by $\tr\tilde\varepsilon=(1-2\nu)g(r)/(1-\nu)$. Subsequently, the refractive index contrast profile is found via Eq.~(\ref{eq:index_contrast}). The corresponding propagation-constant shifts can then be estimated using first-order perturbation theory
  \[
    \delta\beta^{(\ell_1,\ell_2)}=k_0\int\delta n(r)|\psi^{(\ell_1,\ell_2)}|^2\dif A. 
  \]
  The resulting formula is
  \begin{equation}
    \delta \beta^{(\ell_1, \ell_2)}=-C_\beta P \mathcal{M}_{\ell_1}(\zeta) \mathcal{M}_{\ell_2}(\zeta)
    \label{eq:deltabeta}
  \end{equation}
  where
    \[
    \mathcal{M}_{\ell}(\zeta)=
    \frac{1}{\sqrt{\zeta}}\left(\frac{2-\zeta}{\zeta}\right)^{\ell / 2}
    P_{\ell}\left[\frac{1}{\sqrt{\zeta(2-\zeta)}}\right],
  \]
  $P_\ell$ is the respective Legendre polynomial, 
  $C_\beta=k_0n_0^3 p_e\Delta(1+\nu)(1-2 \nu)/(2\pi c_0 R^2E)$,  $\zeta=1+w_0^2/\tilde{R}^2$, and $c_0$ is the speed of light in vacuum.

%%%%

% \bibliography{/Users/nikos/gd/home/active/documents/bibliography/mybib}% Produces the bibliography via BibTeX.

\begin{thebibliography}{44}%
\makeatletter
\providecommand \@ifxundefined [1]{%
 \@ifx{#1\undefined}
}%
\providecommand \@ifnum [1]{%
 \ifnum #1\expandafter \@firstoftwo
 \else \expandafter \@secondoftwo
 \fi
}%
\providecommand \@ifx [1]{%
 \ifx #1\expandafter \@firstoftwo
 \else \expandafter \@secondoftwo
 \fi
}%
\providecommand \natexlab [1]{#1}%
\providecommand \enquote  [1]{``#1''}%
\providecommand \bibnamefont  [1]{#1}%
\providecommand \bibfnamefont [1]{#1}%
\providecommand \citenamefont [1]{#1}%
\providecommand \href@noop [0]{\@secondoftwo}%
\providecommand \href [0]{\begingroup \@sanitize@url \@href}%
\providecommand \@href[1]{\@@startlink{#1}\@@href}%
\providecommand \@@href[1]{\endgroup#1\@@endlink}%
\providecommand \@sanitize@url [0]{\catcode `\\12\catcode `\$12\catcode
  `\&12\catcode `\#12\catcode `\^12\catcode `\_12\catcode `\%12\relax}%
\providecommand \@@startlink[1]{}%
\providecommand \@@endlink[0]{}%
\providecommand \url  [0]{\begingroup\@sanitize@url \@url }%
\providecommand \@url [1]{\endgroup\@href {#1}{\urlprefix }}%
\providecommand \urlprefix  [0]{URL }%
\providecommand \Eprint [0]{\href }%
\providecommand \doibase [0]{https://doi.org/}%
\providecommand \selectlanguage [0]{\@gobble}%
\providecommand \bibinfo  [0]{\@secondoftwo}%
\providecommand \bibfield  [0]{\@secondoftwo}%
\providecommand \translation [1]{[#1]}%
\providecommand \BibitemOpen [0]{}%
\providecommand \bibitemStop [0]{}%
\providecommand \bibitemNoStop [0]{.\EOS\space}%
\providecommand \EOS [0]{\spacefactor3000\relax}%
\providecommand \BibitemShut  [1]{\csname bibitem#1\endcsname}%
\let\auto@bib@innerbib\@empty
%</preamble>
\bibitem [{\citenamefont {Wu}\ \emph {et~al.}(2019)\citenamefont {Wu},
  \citenamefont {Hassan},\ and\ \citenamefont {Christodoulides}}]{wu-np2019}%
  \BibitemOpen
  \bibfield  {author} {\bibinfo {author} {\bibfnamefont {F.~O.}\ \bibnamefont
  {Wu}}, \bibinfo {author} {\bibfnamefont {A.~U.}\ \bibnamefont {Hassan}},\
  and\ \bibinfo {author} {\bibfnamefont {D.~N.}\ \bibnamefont
  {Christodoulides}},\ }\bibfield  {title} {\bibinfo {title} {Thermodynamic
  theory of highly multimoded nonlinear optical systems},\ }\href@noop {}
  {\bibfield  {journal} {\bibinfo  {journal} {Nat. Photonics}\ }\textbf
  {\bibinfo {volume} {13}},\ \bibinfo {pages} {776} (\bibinfo {year}
  {2019})}\BibitemShut {NoStop}%
\bibitem [{Note1()}]{Note1}%
  \BibitemOpen
  \bibinfo {note} {Systems without such conservation laws require a different description, e.g., open-system or nonequilibrium thermodynamics, and fall outside the scope of the present theory. In addition, the thermodynamics of a photon gas in equilibrium with matter is a different physical problem: photons can be absorbed and emitted by the material reservoir, the photon number $N_\gamma$ is not conserved, and therefore its thermodynamic conjugate variable (the photon chemical potential) vanishes. In this case the probability is described by a canonical ensemble $\rho\sim\exp(-\beta U)$}\BibitemShut {NoStop}%
\bibitem [{\citenamefont {Parto}\ \emph {et~al.}(2019)\citenamefont {Parto},
  \citenamefont {Wu}, \citenamefont {Jung}, \citenamefont {Makris},\ and\
  \citenamefont {Christodoulides}}]{parto-ol2019}%
  \BibitemOpen
  \bibfield  {author} {\bibinfo {author} {\bibfnamefont {M.}~\bibnamefont
  {Parto}}, \bibinfo {author} {\bibfnamefont {F.~O.}\ \bibnamefont {Wu}},
  \bibinfo {author} {\bibfnamefont {P.~S.}\ \bibnamefont {Jung}}, \bibinfo
  {author} {\bibfnamefont {K.}~\bibnamefont {Makris}},\ and\ \bibinfo {author}
  {\bibfnamefont {D.~N.}\ \bibnamefont {Christodoulides}},\ }\bibfield  {title}
  {\bibinfo {title} {Thermodynamic conditions governing the optical temperature
  and chemical potential in nonlinear highly multimoded photonic systems},\
  }\href {https://doi.org/10.1364/OL.44.003936} {\bibfield  {journal} {\bibinfo
   {journal} {Opt. Lett.}\ }\textbf {\bibinfo {volume} {44}},\ \bibinfo {pages}
  {3936} (\bibinfo {year} {2019})}\BibitemShut {NoStop}%
\bibitem [{\citenamefont {Lopez-Galmiche}\ \emph {et~al.}(2016)\citenamefont
  {Lopez-Galmiche}, \citenamefont {Eznaveh}, \citenamefont {Eftekhar},
  \citenamefont {Lopez}, \citenamefont {Wright}, \citenamefont {Wise},
  \citenamefont {Christodoulides},\ and\ \citenamefont
  {Correa}}]{lopez-ol2016}%
  \BibitemOpen
  \bibfield  {author} {\bibinfo {author} {\bibfnamefont {G.}~\bibnamefont
  {Lopez-Galmiche}}, \bibinfo {author} {\bibfnamefont {Z.~S.}\ \bibnamefont
  {Eznaveh}}, \bibinfo {author} {\bibfnamefont {M.~A.}\ \bibnamefont
  {Eftekhar}}, \bibinfo {author} {\bibfnamefont {J.~A.}\ \bibnamefont {Lopez}},
  \bibinfo {author} {\bibfnamefont {L.~G.}\ \bibnamefont {Wright}}, \bibinfo
  {author} {\bibfnamefont {F.}~\bibnamefont {Wise}}, \bibinfo {author}
  {\bibfnamefont {D.}~\bibnamefont {Christodoulides}},\ and\ \bibinfo {author}
  {\bibfnamefont {R.~A.}\ \bibnamefont {Correa}},\ }\bibfield  {title}
  {\bibinfo {title} {Visible supercontinuum generation in a graded index
  multimode fiber pumped at 1064 nm},\ }\href
  {https://doi.org/10.1364/OL.41.002553} {\bibfield  {journal} {\bibinfo
  {journal} {Opt. Lett.}\ }\textbf {\bibinfo {volume} {41}},\ \bibinfo {pages}
  {2553} (\bibinfo {year} {2016})}\BibitemShut {NoStop}%
\bibitem [{\citenamefont {Liu}\ \emph {et~al.}(2016)\citenamefont {Liu},
  \citenamefont {Wright}, \citenamefont {Christodoulides},\ and\ \citenamefont
  {Wise}}]{liu-ol2016}%
  \BibitemOpen
  \bibfield  {author} {\bibinfo {author} {\bibfnamefont {Z.}~\bibnamefont
  {Liu}}, \bibinfo {author} {\bibfnamefont {L.~G.}\ \bibnamefont {Wright}},
  \bibinfo {author} {\bibfnamefont {D.~N.}\ \bibnamefont {Christodoulides}},\
  and\ \bibinfo {author} {\bibfnamefont {F.~W.}\ \bibnamefont {Wise}},\
  }\bibfield  {title} {\bibinfo {title} {{K}err self-cleaning of
  femtosecond-pulsed beams in graded-index multimode fiber},\ }\href
  {https://doi.org/10.1364/OL.41.003675} {\bibfield  {journal} {\bibinfo
  {journal} {Opt. Lett.}\ }\textbf {\bibinfo {volume} {41}},\ \bibinfo {pages}
  {3675} (\bibinfo {year} {2016})}\BibitemShut {NoStop}%
\bibitem [{\citenamefont {Krupa}\ \emph {et~al.}(2017)\citenamefont {Krupa},
  \citenamefont {Tonello}, \citenamefont {Shalaby}, \citenamefont {Fabert},
  \citenamefont {Barth{\'e}l{\'e}my}, \citenamefont {Millot}, \citenamefont
  {Wabnitz},\ and\ \citenamefont {Couderc}}]{krupa-np2017}%
  \BibitemOpen
  \bibfield  {author} {\bibinfo {author} {\bibfnamefont {K.}~\bibnamefont
  {Krupa}}, \bibinfo {author} {\bibfnamefont {A.}~\bibnamefont {Tonello}},
  \bibinfo {author} {\bibfnamefont {B.}~\bibnamefont {Shalaby}}, \bibinfo
  {author} {\bibfnamefont {M.}~\bibnamefont {Fabert}}, \bibinfo {author}
  {\bibfnamefont {A.}~\bibnamefont {Barth{\'e}l{\'e}my}}, \bibinfo {author}
  {\bibfnamefont {G.}~\bibnamefont {Millot}}, \bibinfo {author} {\bibfnamefont
  {S.}~\bibnamefont {Wabnitz}},\ and\ \bibinfo {author} {\bibfnamefont
  {V.}~\bibnamefont {Couderc}},\ }\bibfield  {title} {\bibinfo {title} {Spatial
  beam self-cleaning in multimode fibres},\ }\href@noop {} {\bibfield
  {journal} {\bibinfo  {journal} {Nat. Photon.}\ }\textbf {\bibinfo {volume}
  {11}},\ \bibinfo {pages} {237} (\bibinfo {year} {2017})}\BibitemShut
  {NoStop}%
\bibitem [{\citenamefont {Pourbeyram}\ \emph {et~al.}(2022)\citenamefont
  {Pourbeyram}, \citenamefont {Sidorenko}, \citenamefont {Wu}, \citenamefont
  {Bender}, \citenamefont {Wright}, \citenamefont {Christodoulides},\ and\
  \citenamefont {Wise}}]{pourb-np2022}%
  \BibitemOpen
  \bibfield  {author} {\bibinfo {author} {\bibfnamefont {H.}~\bibnamefont
  {Pourbeyram}}, \bibinfo {author} {\bibfnamefont {P.}~\bibnamefont
  {Sidorenko}}, \bibinfo {author} {\bibfnamefont {F.~O.}\ \bibnamefont {Wu}},
  \bibinfo {author} {\bibfnamefont {N.}~\bibnamefont {Bender}}, \bibinfo
  {author} {\bibfnamefont {L.}~\bibnamefont {Wright}}, \bibinfo {author}
  {\bibfnamefont {D.~N.}\ \bibnamefont {Christodoulides}},\ and\ \bibinfo
  {author} {\bibfnamefont {F.}~\bibnamefont {Wise}},\ }\bibfield  {title}
  {\bibinfo {title} {Direct observations of thermalization to a rayleigh--jeans
  distribution in multimode optical fibres},\ }\href
  {https://doi.org/10.1038/s41567-022-01579-y} {\bibfield  {journal} {\bibinfo
  {journal} {Nature Physics}\ }\textbf {\bibinfo {volume} {18}},\ \bibinfo
  {pages} {685} (\bibinfo {year} {2022})}\BibitemShut {NoStop}%
\bibitem [{\citenamefont {Mangini}\ \emph {et~al.}(2022)\citenamefont
  {Mangini}, \citenamefont {Gervaziev}, \citenamefont {Ferraro}, \citenamefont
  {Kharenko}, \citenamefont {Zitelli}, \citenamefont {Sun}, \citenamefont
  {Couderc}, \citenamefont {Podivilov}, \citenamefont {Babin},\ and\
  \citenamefont {Wabnitz}}]{mangi-oe2022}%
  \BibitemOpen
  \bibfield  {author} {\bibinfo {author} {\bibfnamefont {F.}~\bibnamefont
  {Mangini}}, \bibinfo {author} {\bibfnamefont {M.}~\bibnamefont {Gervaziev}},
  \bibinfo {author} {\bibfnamefont {M.}~\bibnamefont {Ferraro}}, \bibinfo
  {author} {\bibfnamefont {D.~S.}\ \bibnamefont {Kharenko}}, \bibinfo {author}
  {\bibfnamefont {M.}~\bibnamefont {Zitelli}}, \bibinfo {author} {\bibfnamefont
  {Y.}~\bibnamefont {Sun}}, \bibinfo {author} {\bibfnamefont {V.}~\bibnamefont
  {Couderc}}, \bibinfo {author} {\bibfnamefont {E.~V.}\ \bibnamefont
  {Podivilov}}, \bibinfo {author} {\bibfnamefont {S.~A.}\ \bibnamefont
  {Babin}},\ and\ \bibinfo {author} {\bibfnamefont {S.}~\bibnamefont
  {Wabnitz}},\ }\bibfield  {title} {\bibinfo {title} {Statistical mechanics of
  beam self-cleaning in grin multimode optical fibers},\ }\href
  {https://doi.org/10.1364/OE.449187} {\bibfield  {journal} {\bibinfo
  {journal} {Opt. Express}\ }\textbf {\bibinfo {volume} {30}},\ \bibinfo
  {pages} {10850} (\bibinfo {year} {2022})}\BibitemShut {NoStop}%
\bibitem [{\citenamefont {{Marques Muniz}}\ \emph {et~al.}(2023)\citenamefont
  {{Marques Muniz}}, \citenamefont {Wu}, \citenamefont {Jung}, \citenamefont
  {Khajavikhan}, \citenamefont {Christodoulides},\ and\ \citenamefont
  {Peschel}}]{muniz-science2023}%
  \BibitemOpen
  \bibfield  {author} {\bibinfo {author} {\bibfnamefont {A.~L.}\ \bibnamefont
  {{Marques Muniz}}}, \bibinfo {author} {\bibfnamefont {F.~O.}\ \bibnamefont
  {Wu}}, \bibinfo {author} {\bibfnamefont {P.~S.}\ \bibnamefont {Jung}},
  \bibinfo {author} {\bibfnamefont {M.}~\bibnamefont {Khajavikhan}}, \bibinfo
  {author} {\bibfnamefont {D.~N.}\ \bibnamefont {Christodoulides}},\ and\
  \bibinfo {author} {\bibfnamefont {U.}~\bibnamefont {Peschel}},\ }\bibfield
  {title} {\bibinfo {title} {Observation of photon-photon thermodynamic
  processes under negative optical temperature conditions},\ }\href
  {https://doi.org/10.1126/science.ade6523} {\bibfield  {journal} {\bibinfo
  {journal} {Science}\ }\textbf {\bibinfo {volume} {379}},\ \bibinfo {pages}
  {1019} (\bibinfo {year} {2023})},\ \Eprint
  {https://arxiv.org/abs/https://www.science.org/doi/pdf/10.1126/science.ade6523}
  {https://www.science.org/doi/pdf/10.1126/science.ade6523} \BibitemShut
  {NoStop}%
\bibitem [{\citenamefont {Baudin}\ \emph {et~al.}(2023)\citenamefont {Baudin},
  \citenamefont {Garnier}, \citenamefont {Fusaro}, \citenamefont {Berti},
  \citenamefont {Michel}, \citenamefont {Krupa}, \citenamefont {Millot},\ and\
  \citenamefont {Picozzi}}]{baudin-prl2023}%
  \BibitemOpen
  \bibfield  {author} {\bibinfo {author} {\bibfnamefont {K.}~\bibnamefont
  {Baudin}}, \bibinfo {author} {\bibfnamefont {J.}~\bibnamefont {Garnier}},
  \bibinfo {author} {\bibfnamefont {A.}~\bibnamefont {Fusaro}}, \bibinfo
  {author} {\bibfnamefont {N.}~\bibnamefont {Berti}}, \bibinfo {author}
  {\bibfnamefont {C.}~\bibnamefont {Michel}}, \bibinfo {author} {\bibfnamefont
  {K.}~\bibnamefont {Krupa}}, \bibinfo {author} {\bibfnamefont
  {G.}~\bibnamefont {Millot}},\ and\ \bibinfo {author} {\bibfnamefont
  {A.}~\bibnamefont {Picozzi}},\ }\bibfield  {title} {\bibinfo {title}
  {Observation of light thermalization to negative-temperature rayleigh-jeans
  equilibrium states in multimode optical fibers},\ }\href
  {https://doi.org/10.1103/PhysRevLett.130.063801} {\bibfield  {journal}
  {\bibinfo  {journal} {Phys. Rev. Lett.}\ }\textbf {\bibinfo {volume} {130}},\
  \bibinfo {pages} {063801} (\bibinfo {year} {2023})}\BibitemShut {NoStop}%
\bibitem [{\citenamefont {Wu}\ \emph {et~al.}(2022)\citenamefont {Wu},
  \citenamefont {Zhong}, \citenamefont {Ren}, \citenamefont {Jung},
  \citenamefont {Makris},\ and\ \citenamefont {Christodoulides}}]{wu-prl2022}%
  \BibitemOpen
  \bibfield  {author} {\bibinfo {author} {\bibfnamefont {F.~O.}\ \bibnamefont
  {Wu}}, \bibinfo {author} {\bibfnamefont {Q.}~\bibnamefont {Zhong}}, \bibinfo
  {author} {\bibfnamefont {H.}~\bibnamefont {Ren}}, \bibinfo {author}
  {\bibfnamefont {P.~S.}\ \bibnamefont {Jung}}, \bibinfo {author}
  {\bibfnamefont {K.~G.}\ \bibnamefont {Makris}},\ and\ \bibinfo {author}
  {\bibfnamefont {D.~N.}\ \bibnamefont {Christodoulides}},\ }\bibfield  {title}
  {\bibinfo {title} {Thermalization of light's orbital angular momentum in
  nonlinear multimode waveguide systems},\ }\href
  {https://doi.org/10.1103/PhysRevLett.128.123901} {\bibfield  {journal}
  {\bibinfo  {journal} {Phys. Rev. Lett.}\ }\textbf {\bibinfo {volume} {128}},\
  \bibinfo {pages} {123901} (\bibinfo {year} {2022})}\BibitemShut {NoStop}%
\bibitem [{\citenamefont {Podivilov}\ \emph {et~al.}(2022)\citenamefont
  {Podivilov}, \citenamefont {Mangini}, \citenamefont {Sidelnikov},
  \citenamefont {Ferraro}, \citenamefont {Gervaziev}, \citenamefont {Kharenko},
  \citenamefont {Zitelli}, \citenamefont {Fedoruk}, \citenamefont {Babin},\
  and\ \citenamefont {Wabnitz}}]{podiv-prl2022}%
  \BibitemOpen
  \bibfield  {author} {\bibinfo {author} {\bibfnamefont {E.~V.}\ \bibnamefont
  {Podivilov}}, \bibinfo {author} {\bibfnamefont {F.}~\bibnamefont {Mangini}},
  \bibinfo {author} {\bibfnamefont {O.~S.}\ \bibnamefont {Sidelnikov}},
  \bibinfo {author} {\bibfnamefont {M.}~\bibnamefont {Ferraro}}, \bibinfo
  {author} {\bibfnamefont {M.}~\bibnamefont {Gervaziev}}, \bibinfo {author}
  {\bibfnamefont {D.~S.}\ \bibnamefont {Kharenko}}, \bibinfo {author}
  {\bibfnamefont {M.}~\bibnamefont {Zitelli}}, \bibinfo {author} {\bibfnamefont
  {M.~P.}\ \bibnamefont {Fedoruk}}, \bibinfo {author} {\bibfnamefont {S.~A.}\
  \bibnamefont {Babin}},\ and\ \bibinfo {author} {\bibfnamefont
  {S.}~\bibnamefont {Wabnitz}},\ }\bibfield  {title} {\bibinfo {title}
  {Thermalization of orbital angular momentum beams in multimode optical
  fibers},\ }\href {https://doi.org/10.1103/PhysRevLett.128.243901} {\bibfield
  {journal} {\bibinfo  {journal} {Phys. Rev. Lett.}\ }\textbf {\bibinfo
  {volume} {128}},\ \bibinfo {pages} {243901} (\bibinfo {year}
  {2022})}\BibitemShut {NoStop}%
\bibitem [{\citenamefont {Kirsch}\ \emph {et~al.}(2025)\citenamefont {Kirsch},
  \citenamefont {Pyrialakos}, \citenamefont {Altenkirch}, \citenamefont
  {A.~Selim}, \citenamefont {Beck}, \citenamefont {Wolterink}, \citenamefont
  {Ren}, \citenamefont {Jung}, \citenamefont {Khajavikhan}, \citenamefont
  {Szameit} \emph {et~al.}}]{kirsc-np2025}%
  \BibitemOpen
  \bibfield  {author} {\bibinfo {author} {\bibfnamefont {M.~S.}\ \bibnamefont
  {Kirsch}}, \bibinfo {author} {\bibfnamefont {G.~G.}\ \bibnamefont
  {Pyrialakos}}, \bibinfo {author} {\bibfnamefont {R.}~\bibnamefont
  {Altenkirch}}, \bibinfo {author} {\bibfnamefont {M.}~\bibnamefont
  {A.~Selim}}, \bibinfo {author} {\bibfnamefont {J.}~\bibnamefont {Beck}},
  \bibinfo {author} {\bibfnamefont {T.~A.}\ \bibnamefont {Wolterink}}, \bibinfo
  {author} {\bibfnamefont {H.}~\bibnamefont {Ren}}, \bibinfo {author}
  {\bibfnamefont {P.~S.}\ \bibnamefont {Jung}}, \bibinfo {author}
  {\bibfnamefont {M.}~\bibnamefont {Khajavikhan}}, \bibinfo {author}
  {\bibfnamefont {A.}~\bibnamefont {Szameit}}, \emph {et~al.},\ }\bibfield
  {title} {\bibinfo {title} {Observation of joule--thomson photon-gas
  expansion},\ }\href@noop {} {\bibfield  {journal} {\bibinfo  {journal}
  {Nature Physics}\ }\textbf {\bibinfo {volume} {21}},\ \bibinfo {pages} {214}
  (\bibinfo {year} {2025})}\BibitemShut {NoStop}%
\bibitem [{\citenamefont {Makris}\ \emph {et~al.}(2020)\citenamefont {Makris},
  \citenamefont {Wu}, \citenamefont {Jung},\ and\ \citenamefont
  {Christodoulides}}]{makri-ol2020}%
  \BibitemOpen
  \bibfield  {author} {\bibinfo {author} {\bibfnamefont {K.~G.}\ \bibnamefont
  {Makris}}, \bibinfo {author} {\bibfnamefont {F.~O.}\ \bibnamefont {Wu}},
  \bibinfo {author} {\bibfnamefont {P.~S.}\ \bibnamefont {Jung}},\ and\
  \bibinfo {author} {\bibfnamefont {D.~N.}\ \bibnamefont {Christodoulides}},\
  }\bibfield  {title} {\bibinfo {title} {Statistical mechanics of weakly
  nonlinear optical multimode gases},\ }\href
  {https://doi.org/10.1364/OL.387863} {\bibfield  {journal} {\bibinfo
  {journal} {Opt. Lett.}\ }\textbf {\bibinfo {volume} {45}},\ \bibinfo {pages}
  {1651} (\bibinfo {year} {2020})}\BibitemShut {NoStop}%
\bibitem [{\citenamefont {Efremidis}\ and\ \citenamefont
  {Christodoulides}(2021)}]{efrem-pra2021}%
  \BibitemOpen
  \bibfield  {author} {\bibinfo {author} {\bibfnamefont {N.~K.}\ \bibnamefont
  {Efremidis}}\ and\ \bibinfo {author} {\bibfnamefont {D.~N.}\ \bibnamefont
  {Christodoulides}},\ }\bibfield  {title} {\bibinfo {title} {Fundamental
  entropic processes in the theory of optical thermodynamics},\ }\href
  {https://doi.org/10.1103/PhysRevA.103.043517} {\bibfield  {journal} {\bibinfo
   {journal} {Phys. Rev. A}\ }\textbf {\bibinfo {volume} {103}},\ \bibinfo
  {pages} {043517} (\bibinfo {year} {2021})}\BibitemShut {NoStop}%
\bibitem [{\citenamefont {Jung}\ \emph {et~al.}(2022)\citenamefont {Jung},
  \citenamefont {Pyrialakos}, \citenamefont {Wu}, \citenamefont {Parto},
  \citenamefont {Khajavikhan}, \citenamefont {Krolikowski},\ and\ \citenamefont
  {Christodoulides}}]{jung-nc2022}%
  \BibitemOpen
  \bibfield  {author} {\bibinfo {author} {\bibfnamefont {P.~S.}\ \bibnamefont
  {Jung}}, \bibinfo {author} {\bibfnamefont {G.~G.}\ \bibnamefont
  {Pyrialakos}}, \bibinfo {author} {\bibfnamefont {F.~O.}\ \bibnamefont {Wu}},
  \bibinfo {author} {\bibfnamefont {M.}~\bibnamefont {Parto}}, \bibinfo
  {author} {\bibfnamefont {M.}~\bibnamefont {Khajavikhan}}, \bibinfo {author}
  {\bibfnamefont {W.}~\bibnamefont {Krolikowski}},\ and\ \bibinfo {author}
  {\bibfnamefont {D.~N.}\ \bibnamefont {Christodoulides}},\ }\bibfield  {title}
  {\bibinfo {title} {Thermal control of the topological edge flow in nonlinear
  photonic lattices},\ }\href {https://doi.org/10.1038/s41467-022-32069-7}
  {\bibfield  {journal} {\bibinfo  {journal} {Nature Communications}\ }\textbf
  {\bibinfo {volume} {13}},\ \bibinfo {pages} {4393} (\bibinfo {year}
  {2022})}\BibitemShut {NoStop}%
\bibitem [{\citenamefont {Berti}\ \emph {et~al.}(2022)\citenamefont {Berti},
  \citenamefont {Baudin}, \citenamefont {Fusaro}, \citenamefont {Millot},
  \citenamefont {Picozzi},\ and\ \citenamefont {Garnier}}]{berti-prl2022}%
  \BibitemOpen
  \bibfield  {author} {\bibinfo {author} {\bibfnamefont {N.}~\bibnamefont
  {Berti}}, \bibinfo {author} {\bibfnamefont {K.}~\bibnamefont {Baudin}},
  \bibinfo {author} {\bibfnamefont {A.}~\bibnamefont {Fusaro}}, \bibinfo
  {author} {\bibfnamefont {G.}~\bibnamefont {Millot}}, \bibinfo {author}
  {\bibfnamefont {A.}~\bibnamefont {Picozzi}},\ and\ \bibinfo {author}
  {\bibfnamefont {J.}~\bibnamefont {Garnier}},\ }\bibfield  {title} {\bibinfo
  {title} {Interplay of thermalization and strong disorder: Wave turbulence
  theory, numerical simulations, and experiments in multimode optical fibers},\
  }\href {https://doi.org/10.1103/PhysRevLett.129.063901} {\bibfield  {journal}
  {\bibinfo  {journal} {Phys. Rev. Lett.}\ }\textbf {\bibinfo {volume} {129}},\
  \bibinfo {pages} {063901} (\bibinfo {year} {2022})}\BibitemShut {NoStop}%
\bibitem [{\citenamefont {Ramos}\ \emph {et~al.}(2023)\citenamefont {Ramos},
  \citenamefont {Shi}, \citenamefont {Fernández-Alcázar}, \citenamefont
  {Christodoulides},\ and\ \citenamefont {Kottos}}]{ramos-cp2023}%
  \BibitemOpen
  \bibfield  {author} {\bibinfo {author} {\bibfnamefont {A.~Y.}\ \bibnamefont
  {Ramos}}, \bibinfo {author} {\bibfnamefont {C.}~\bibnamefont {Shi}}, \bibinfo
  {author} {\bibfnamefont {L.~J.}\ \bibnamefont {Fernández-Alcázar}},
  \bibinfo {author} {\bibfnamefont {D.~N.}\ \bibnamefont {Christodoulides}},\
  and\ \bibinfo {author} {\bibfnamefont {T.}~\bibnamefont {Kottos}},\
  }\bibfield  {title} {\bibinfo {title} {Theory of localization-hindered
  thermalization in nonlinear multimode photonics},\ }\href
  {https://doi.org/10.1038/s42005-023-01309-7} {\bibfield  {journal} {\bibinfo
  {journal} {Communications Physics}\ }\textbf {\bibinfo {volume} {6}},\
  \bibinfo {pages} {189} (\bibinfo {year} {2023})}\BibitemShut {NoStop}%
\bibitem [{\citenamefont {Yang}\ \emph {et~al.}(2024)\citenamefont {Yang},
  \citenamefont {Bongiovanni}, \citenamefont {Song}, \citenamefont
  {Morandotti}, \citenamefont {Chen},\ and\ \citenamefont
  {Efremidis}}]{yang-aplp2024}%
  \BibitemOpen
  \bibfield  {author} {\bibinfo {author} {\bibfnamefont {G.}~\bibnamefont
  {Yang}}, \bibinfo {author} {\bibfnamefont {D.}~\bibnamefont {Bongiovanni}},
  \bibinfo {author} {\bibfnamefont {D.}~\bibnamefont {Song}}, \bibinfo {author}
  {\bibfnamefont {R.}~\bibnamefont {Morandotti}}, \bibinfo {author}
  {\bibfnamefont {Z.}~\bibnamefont {Chen}},\ and\ \bibinfo {author}
  {\bibfnamefont {N.~K.}\ \bibnamefont {Efremidis}},\ }\bibfield  {title}
  {\bibinfo {title} {Thermalization dynamics in photonic lattices of different
  geometries},\ }\href {https://doi.org/10.1063/5.0205202} {\bibfield
  {journal} {\bibinfo  {journal} {APL Photonics}\ }\textbf {\bibinfo {volume}
  {9}},\ \bibinfo {pages} {066115} (\bibinfo {year} {2024})},\ \Eprint
  {https://arxiv.org/abs/https://pubs.aip.org/aip/app/article-pdf/doi/10.1063/5.0205202/20008776/066115\_1\_5.0205202.pdf}
  {https://pubs.aip.org/aip/app/article-pdf/doi/10.1063/5.0205202/20008776/066115\_1\_5.0205202.pdf}
  \BibitemShut {NoStop}%
\bibitem [{\citenamefont {Yang}\ \emph {et~al.}(2025)\citenamefont {Yang},
  \citenamefont {Wang}, \citenamefont {Chen}, \citenamefont {Song},
  \citenamefont {Xia}, \citenamefont {Song}, \citenamefont {Chen},\ and\
  \citenamefont {Efremidis}}]{yang-aplp2025}%
  \BibitemOpen
  \bibfield  {author} {\bibinfo {author} {\bibfnamefont {G.}~\bibnamefont
  {Yang}}, \bibinfo {author} {\bibfnamefont {J.}~\bibnamefont {Wang}}, \bibinfo
  {author} {\bibfnamefont {Y.}~\bibnamefont {Chen}}, \bibinfo {author}
  {\bibfnamefont {L.}~\bibnamefont {Song}}, \bibinfo {author} {\bibfnamefont
  {S.}~\bibnamefont {Xia}}, \bibinfo {author} {\bibfnamefont {D.}~\bibnamefont
  {Song}}, \bibinfo {author} {\bibfnamefont {Z.}~\bibnamefont {Chen}},\ and\
  \bibinfo {author} {\bibfnamefont {N.~K.}\ \bibnamefont {Efremidis}},\
  }\bibfield  {title} {\bibinfo {title} {Unveiling prethermalization and
  thermal processes through the simplest one-dimensional topological model},\
  }\href {https://doi.org/10.1063/5.0270636} {\bibfield  {journal} {\bibinfo
  {journal} {APL Photonics}\ }\textbf {\bibinfo {volume} {10}},\ \bibinfo
  {pages} {076114} (\bibinfo {year} {2025})},\ \Eprint
  {https://arxiv.org/abs/https://pubs.aip.org/aip/app/article-pdf/doi/10.1063/5.0270636/20601006/076114\_1\_5.0270636.pdf}
  {https://pubs.aip.org/aip/app/article-pdf/doi/10.1063/5.0270636/20601006/076114\_1\_5.0270636.pdf}
  \BibitemShut {NoStop}%
\bibitem [{\citenamefont {Ferraro}\ \emph {et~al.}(2025)\citenamefont
  {Ferraro}, \citenamefont {Baudin}, \citenamefont {Gervaziev}, \citenamefont
  {Fusaro}, \citenamefont {Picozzi}, \citenamefont {Garnier}, \citenamefont
  {Millot}, \citenamefont {Kharenko}, \citenamefont {Podivilov}, \citenamefont
  {Babin}, \citenamefont {Mangini},\ and\ \citenamefont
  {Wabnitz}}]{ferra-PhysD2025}%
  \BibitemOpen
  \bibfield  {author} {\bibinfo {author} {\bibfnamefont {M.}~\bibnamefont
  {Ferraro}}, \bibinfo {author} {\bibfnamefont {K.}~\bibnamefont {Baudin}},
  \bibinfo {author} {\bibfnamefont {M.}~\bibnamefont {Gervaziev}}, \bibinfo
  {author} {\bibfnamefont {A.}~\bibnamefont {Fusaro}}, \bibinfo {author}
  {\bibfnamefont {A.}~\bibnamefont {Picozzi}}, \bibinfo {author} {\bibfnamefont
  {J.}~\bibnamefont {Garnier}}, \bibinfo {author} {\bibfnamefont
  {G.}~\bibnamefont {Millot}}, \bibinfo {author} {\bibfnamefont
  {D.}~\bibnamefont {Kharenko}}, \bibinfo {author} {\bibfnamefont
  {E.}~\bibnamefont {Podivilov}}, \bibinfo {author} {\bibfnamefont
  {S.}~\bibnamefont {Babin}}, \bibinfo {author} {\bibfnamefont
  {F.}~\bibnamefont {Mangini}},\ and\ \bibinfo {author} {\bibfnamefont
  {S.}~\bibnamefont {Wabnitz}},\ }\bibfield  {title} {\bibinfo {title} {Wave
  turbulence, thermalization and multimode locking in optical fibers},\ }\href
  {https://doi.org/https://doi.org/10.1016/j.physd.2025.134758} {\bibfield
  {journal} {\bibinfo  {journal} {Physica D: Nonlinear Phenomena}\ }\textbf
  {\bibinfo {volume} {481}},\ \bibinfo {pages} {134758} (\bibinfo {year}
  {2025})}\BibitemShut {NoStop}%
\bibitem [{\citenamefont {Povinelli}\ \emph
  {et~al.}(2005{\natexlab{a}})\citenamefont {Povinelli}, \citenamefont
  {Lon\v{c}ar}, \citenamefont {Ibanescu}, \citenamefont {Smythe}, \citenamefont
  {Johnson}, \citenamefont {Capasso},\ and\ \citenamefont
  {Joannopoulos}}]{povin-ol2005}%
  \BibitemOpen
  \bibfield  {author} {\bibinfo {author} {\bibfnamefont {M.~L.}\ \bibnamefont
  {Povinelli}}, \bibinfo {author} {\bibfnamefont {M.}~\bibnamefont
  {Lon\v{c}ar}}, \bibinfo {author} {\bibfnamefont {M.}~\bibnamefont
  {Ibanescu}}, \bibinfo {author} {\bibfnamefont {E.~J.}\ \bibnamefont
  {Smythe}}, \bibinfo {author} {\bibfnamefont {S.~G.}\ \bibnamefont {Johnson}},
  \bibinfo {author} {\bibfnamefont {F.}~\bibnamefont {Capasso}},\ and\ \bibinfo
  {author} {\bibfnamefont {J.~D.}\ \bibnamefont {Joannopoulos}},\ }\bibfield
  {title} {\bibinfo {title} {Evanescent-wave bonding between optical
  waveguides},\ }\href {https://doi.org/10.1364/OL.30.003042} {\bibfield
  {journal} {\bibinfo  {journal} {Opt. Lett.}\ }\textbf {\bibinfo {volume}
  {30}},\ \bibinfo {pages} {3042} (\bibinfo {year}
  {2005}{\natexlab{a}})}\BibitemShut {NoStop}%
\bibitem [{\citenamefont {Povinelli}\ \emph
  {et~al.}(2005{\natexlab{b}})\citenamefont {Povinelli}, \citenamefont
  {Johnson}, \citenamefont {Lon\v{c}ar}, \citenamefont {Ibanescu},
  \citenamefont {Smythe}, \citenamefont {Capasso},\ and\ \citenamefont
  {Joannopoulos}}]{povin-oe2005}%
  \BibitemOpen
  \bibfield  {author} {\bibinfo {author} {\bibfnamefont {M.~L.}\ \bibnamefont
  {Povinelli}}, \bibinfo {author} {\bibfnamefont {S.~G.}\ \bibnamefont
  {Johnson}}, \bibinfo {author} {\bibfnamefont {M.}~\bibnamefont {Lon\v{c}ar}},
  \bibinfo {author} {\bibfnamefont {M.}~\bibnamefont {Ibanescu}}, \bibinfo
  {author} {\bibfnamefont {E.~J.}\ \bibnamefont {Smythe}}, \bibinfo {author}
  {\bibfnamefont {F.}~\bibnamefont {Capasso}},\ and\ \bibinfo {author}
  {\bibfnamefont {J.~D.}\ \bibnamefont {Joannopoulos}},\ }\bibfield  {title}
  {\bibinfo {title} {High-{Q} enhancement of attractive and repulsive optical
  forces between coupled whispering-gallery-mode resonators},\ }\href
  {https://doi.org/10.1364/OPEX.13.008286} {\bibfield  {journal} {\bibinfo
  {journal} {Opt. Express}\ }\textbf {\bibinfo {volume} {13}},\ \bibinfo
  {pages} {8286} (\bibinfo {year} {2005}{\natexlab{b}})}\BibitemShut {NoStop}%
\bibitem [{\citenamefont {Rakich}\ \emph {et~al.}(2009)\citenamefont {Rakich},
  \citenamefont {Popovi\'{c}},\ and\ \citenamefont {Wang}}]{rakic-oe2009}%
  \BibitemOpen
  \bibfield  {author} {\bibinfo {author} {\bibfnamefont {P.~T.}\ \bibnamefont
  {Rakich}}, \bibinfo {author} {\bibfnamefont {M.~A.}\ \bibnamefont
  {Popovi\'{c}}},\ and\ \bibinfo {author} {\bibfnamefont {Z.}~\bibnamefont
  {Wang}},\ }\bibfield  {title} {\bibinfo {title} {General treatment of optical
  forces and potentials in mechanically variable photonic systems},\ }\href
  {https://doi.org/10.1364/OE.17.018116} {\bibfield  {journal} {\bibinfo
  {journal} {Opt. Express}\ }\textbf {\bibinfo {volume} {17}},\ \bibinfo
  {pages} {18116} (\bibinfo {year} {2009})}\BibitemShut {NoStop}%
\bibitem [{\citenamefont {Rakich}\ \emph {et~al.}(2011)\citenamefont {Rakich},
  \citenamefont {Wang},\ and\ \citenamefont {Davids}}]{rakic-ol2011}%
  \BibitemOpen
  \bibfield  {author} {\bibinfo {author} {\bibfnamefont {P.~T.}\ \bibnamefont
  {Rakich}}, \bibinfo {author} {\bibfnamefont {Z.}~\bibnamefont {Wang}},\ and\
  \bibinfo {author} {\bibfnamefont {P.}~\bibnamefont {Davids}},\ }\bibfield
  {title} {\bibinfo {title} {Scaling of optical forces in dielectric
  waveguides: rigorous connection between radiation pressure and dispersion},\
  }\href {https://doi.org/10.1364/OL.36.000217} {\bibfield  {journal} {\bibinfo
   {journal} {Opt. Lett.}\ }\textbf {\bibinfo {volume} {36}},\ \bibinfo {pages}
  {217} (\bibinfo {year} {2011})}\BibitemShut {NoStop}%
\bibitem [{\citenamefont {Ren}\ \emph {et~al.}(2022)\citenamefont {Ren},
  \citenamefont {Luo}, \citenamefont {Selim}, \citenamefont {Pyrialakos},
  \citenamefont {Wu}, \citenamefont {Khajavikhan},\ and\ \citenamefont
  {Christodoulides}}]{ren-pra2022}%
  \BibitemOpen
  \bibfield  {author} {\bibinfo {author} {\bibfnamefont {H.}~\bibnamefont
  {Ren}}, \bibinfo {author} {\bibfnamefont {H.}~\bibnamefont {Luo}}, \bibinfo
  {author} {\bibfnamefont {M.~A.}\ \bibnamefont {Selim}}, \bibinfo {author}
  {\bibfnamefont {G.~G.}\ \bibnamefont {Pyrialakos}}, \bibinfo {author}
  {\bibfnamefont {F.~O.}\ \bibnamefont {Wu}}, \bibinfo {author} {\bibfnamefont
  {M.}~\bibnamefont {Khajavikhan}},\ and\ \bibinfo {author} {\bibfnamefont
  {D.}~\bibnamefont {Christodoulides}},\ }\bibfield  {title} {\bibinfo {title}
  {Rigorous analysis of optical forces in dielectric structures based on the
  minkowski-helmholtz formula},\ }\href
  {https://doi.org/10.1103/PhysRevA.106.033517} {\bibfield  {journal} {\bibinfo
   {journal} {Phys. Rev. A}\ }\textbf {\bibinfo {volume} {106}},\ \bibinfo
  {pages} {033517} (\bibinfo {year} {2022})}\BibitemShut {NoStop}%
\bibitem [{\citenamefont {Li}\ \emph {et~al.}(2008)\citenamefont {Li},
  \citenamefont {Pernice}, \citenamefont {Xiong}, \citenamefont {Baehr-Jones},
  \citenamefont {Hochberg},\ and\ \citenamefont {Tang}}]{li-nature2008}%
  \BibitemOpen
  \bibfield  {author} {\bibinfo {author} {\bibfnamefont {M.}~\bibnamefont
  {Li}}, \bibinfo {author} {\bibfnamefont {W.}~\bibnamefont {Pernice}},
  \bibinfo {author} {\bibfnamefont {C.}~\bibnamefont {Xiong}}, \bibinfo
  {author} {\bibfnamefont {T.}~\bibnamefont {Baehr-Jones}}, \bibinfo {author}
  {\bibfnamefont {M.}~\bibnamefont {Hochberg}},\ and\ \bibinfo {author}
  {\bibfnamefont {H.}~\bibnamefont {Tang}},\ }\bibfield  {title} {\bibinfo
  {title} {Harnessing optical forces in integrated photonic circuits},\
  }\href@noop {} {\bibfield  {journal} {\bibinfo  {journal} {Nature}\ }\textbf
  {\bibinfo {volume} {456}},\ \bibinfo {pages} {480} (\bibinfo {year}
  {2008})}\BibitemShut {NoStop}%
\bibitem [{\citenamefont {Li}\ \emph {et~al.}(2009)\citenamefont {Li},
  \citenamefont {Pernice},\ and\ \citenamefont {Tang}}]{li-np2009}%
  \BibitemOpen
  \bibfield  {author} {\bibinfo {author} {\bibfnamefont {M.}~\bibnamefont
  {Li}}, \bibinfo {author} {\bibfnamefont {W.~H.~P.}\ \bibnamefont {Pernice}},\
  and\ \bibinfo {author} {\bibfnamefont {H.~X.}\ \bibnamefont {Tang}},\
  }\bibfield  {title} {\bibinfo {title} {Tunable bipolar optical interactions
  between guided lightwaves},\ }\href
  {https://doi.org/10.1038/nphoton.2009.116} {\bibfield  {journal} {\bibinfo
  {journal} {Nature Photonics}\ }\textbf {\bibinfo {volume} {3}},\ \bibinfo
  {pages} {464} (\bibinfo {year} {2009})}\BibitemShut {NoStop}%
\bibitem [{\citenamefont {Roels}\ \emph {et~al.}(2009)\citenamefont {Roels},
  \citenamefont {De~Vlaminck}, \citenamefont {Lagae}, \citenamefont {Maes},
  \citenamefont {Van~Thourhout},\ and\ \citenamefont {Baets}}]{roels-nn2009}%
  \BibitemOpen
  \bibfield  {author} {\bibinfo {author} {\bibfnamefont {J.}~\bibnamefont
  {Roels}}, \bibinfo {author} {\bibfnamefont {I.}~\bibnamefont {De~Vlaminck}},
  \bibinfo {author} {\bibfnamefont {L.}~\bibnamefont {Lagae}}, \bibinfo
  {author} {\bibfnamefont {B.}~\bibnamefont {Maes}}, \bibinfo {author}
  {\bibfnamefont {D.}~\bibnamefont {Van~Thourhout}},\ and\ \bibinfo {author}
  {\bibfnamefont {R.}~\bibnamefont {Baets}},\ }\bibfield  {title} {\bibinfo
  {title} {Tunable optical forces between nanophotonic waveguides},\ }\href
  {https://doi.org/10.1038/nnano.2009.186} {\bibfield  {journal} {\bibinfo
  {journal} {Nature Nanotechnology}\ }\textbf {\bibinfo {volume} {4}},\
  \bibinfo {pages} {510} (\bibinfo {year} {2009})}\BibitemShut {NoStop}%
\bibitem [{\citenamefont {Van~Thourhout}\ and\ \citenamefont
  {Roels}(2010)}]{vanth-np2010}%
  \BibitemOpen
  \bibfield  {author} {\bibinfo {author} {\bibfnamefont {D.}~\bibnamefont
  {Van~Thourhout}}\ and\ \bibinfo {author} {\bibfnamefont {J.}~\bibnamefont
  {Roels}},\ }\bibfield  {title} {\bibinfo {title} {Optomechanical device
  actuation through the optical gradient force},\ }\href
  {https://doi.org/10.1038/nphoton.2010.72} {\bibfield  {journal} {\bibinfo
  {journal} {Nature Photonics}\ }\textbf {\bibinfo {volume} {4}},\ \bibinfo
  {pages} {211} (\bibinfo {year} {2010})}\BibitemShut {NoStop}%
\bibitem [{\citenamefont {Metcalfe}(2014)}]{metac-apr2014}%
  \BibitemOpen
  \bibfield  {author} {\bibinfo {author} {\bibfnamefont {M.}~\bibnamefont
  {Metcalfe}},\ }\bibfield  {title} {\bibinfo {title} {Applications of cavity
  optomechanics},\ }\href {https://doi.org/10.1063/1.4896029} {\bibfield
  {journal} {\bibinfo  {journal} {Applied Physics Reviews}\ }\textbf {\bibinfo
  {volume} {1}},\ \bibinfo {pages} {031105} (\bibinfo {year} {2014})},\ \Eprint
  {https://arxiv.org/abs/https://doi.org/10.1063/1.4896029}
  {https://doi.org/10.1063/1.4896029} \BibitemShut {NoStop}%
\bibitem [{\citenamefont {Rosenberg}\ \emph {et~al.}(2009)\citenamefont
  {Rosenberg}, \citenamefont {Lin},\ and\ \citenamefont
  {Painter}}]{rosen-np2009}%
  \BibitemOpen
  \bibfield  {author} {\bibinfo {author} {\bibfnamefont {J.}~\bibnamefont
  {Rosenberg}}, \bibinfo {author} {\bibfnamefont {Q.}~\bibnamefont {Lin}},\
  and\ \bibinfo {author} {\bibfnamefont {O.}~\bibnamefont {Painter}},\
  }\bibfield  {title} {\bibinfo {title} {Static and dynamic wavelength routing
  via the gradient optical force},\ }\href
  {https://doi.org/10.1038/nphoton.2009.137} {\bibfield  {journal} {\bibinfo
  {journal} {Nature Photonics}\ }\textbf {\bibinfo {volume} {3}},\ \bibinfo
  {pages} {478} (\bibinfo {year} {2009})}\BibitemShut {NoStop}%
\bibitem [{\citenamefont {Huang}\ \emph {et~al.}(2019)\citenamefont {Huang},
  \citenamefont {Chin}, \citenamefont {Cai}, \citenamefont {Li}, \citenamefont
  {Wu}, \citenamefont {Chen}, \citenamefont {Li},\ and\ \citenamefont
  {Liu}}]{huang-acs2019}%
  \BibitemOpen
  \bibfield  {author} {\bibinfo {author} {\bibfnamefont {J.}~\bibnamefont
  {Huang}}, \bibinfo {author} {\bibfnamefont {L.~K.}\ \bibnamefont {Chin}},
  \bibinfo {author} {\bibfnamefont {H.}~\bibnamefont {Cai}}, \bibinfo {author}
  {\bibfnamefont {H.}~\bibnamefont {Li}}, \bibinfo {author} {\bibfnamefont
  {J.~H.}\ \bibnamefont {Wu}}, \bibinfo {author} {\bibfnamefont
  {T.}~\bibnamefont {Chen}}, \bibinfo {author} {\bibfnamefont {M.}~\bibnamefont
  {Li}},\ and\ \bibinfo {author} {\bibfnamefont {A.-Q.}\ \bibnamefont {Liu}},\
  }\bibfield  {title} {\bibinfo {title} {Dynamic phonon manipulation by
  optomechanically induced strong coupling between two distinct mechanical
  resonators},\ }\href {https://doi.org/10.1021/acsphotonics.9b00618}
  {\bibfield  {journal} {\bibinfo  {journal} {ACS Photonics}\ }\textbf
  {\bibinfo {volume} {6}},\ \bibinfo {pages} {1855} (\bibinfo {year}
  {2019})}\BibitemShut {NoStop}%
\bibitem [{\citenamefont {Anetsberger}\ \emph {et~al.}(2009)\citenamefont
  {Anetsberger}, \citenamefont {Arcizet}, \citenamefont {Unterreithmeier},
  \citenamefont {Rivi{\`e}re}, \citenamefont {Schliesser}, \citenamefont
  {Weig}, \citenamefont {Kotthaus},\ and\ \citenamefont
  {Kippenberg}}]{anets-np2009}%
  \BibitemOpen
  \bibfield  {author} {\bibinfo {author} {\bibfnamefont {G.}~\bibnamefont
  {Anetsberger}}, \bibinfo {author} {\bibfnamefont {O.}~\bibnamefont
  {Arcizet}}, \bibinfo {author} {\bibfnamefont {Q.~P.}\ \bibnamefont
  {Unterreithmeier}}, \bibinfo {author} {\bibfnamefont {R.}~\bibnamefont
  {Rivi{\`e}re}}, \bibinfo {author} {\bibfnamefont {A.}~\bibnamefont
  {Schliesser}}, \bibinfo {author} {\bibfnamefont {E.~M.}\ \bibnamefont
  {Weig}}, \bibinfo {author} {\bibfnamefont {J.~P.}\ \bibnamefont {Kotthaus}},\
  and\ \bibinfo {author} {\bibfnamefont {T.~J.}\ \bibnamefont {Kippenberg}},\
  }\bibfield  {title} {\bibinfo {title} {Near-field cavity optomechanics with
  nanomechanical oscillators},\ }\href {https://doi.org/10.1038/nphys1425}
  {\bibfield  {journal} {\bibinfo  {journal} {Nature Physics}\ }\textbf
  {\bibinfo {volume} {5}},\ \bibinfo {pages} {909} (\bibinfo {year}
  {2009})}\BibitemShut {NoStop}%
\bibitem [{\citenamefont {Pruessner}\ \emph {et~al.}(2018)\citenamefont
  {Pruessner}, \citenamefont {Park}, \citenamefont {Stievater}, \citenamefont
  {Kozak},\ and\ \citenamefont {Rabinovich}}]{prues-acs2018}%
  \BibitemOpen
  \bibfield  {author} {\bibinfo {author} {\bibfnamefont {M.~W.}\ \bibnamefont
  {Pruessner}}, \bibinfo {author} {\bibfnamefont {D.}~\bibnamefont {Park}},
  \bibinfo {author} {\bibfnamefont {T.~H.}\ \bibnamefont {Stievater}}, \bibinfo
  {author} {\bibfnamefont {D.~A.}\ \bibnamefont {Kozak}},\ and\ \bibinfo
  {author} {\bibfnamefont {W.~S.}\ \bibnamefont {Rabinovich}},\ }\bibfield
  {title} {\bibinfo {title} {Optomechanical cavities for all-optical
  photothermal sensing},\ }\href {https://doi.org/10.1021/acsphotonics.8b00452}
  {\bibfield  {journal} {\bibinfo  {journal} {ACS Photonics}\ }\textbf
  {\bibinfo {volume} {5}},\ \bibinfo {pages} {3214} (\bibinfo {year}
  {2018})}\BibitemShut {NoStop}%
\bibitem [{\citenamefont {Ren}\ \emph {et~al.}(2013)\citenamefont {Ren},
  \citenamefont {Huang}, \citenamefont {Cai}, \citenamefont {Tsai},
  \citenamefont {Zhou}, \citenamefont {Liu}, \citenamefont {Suo},\ and\
  \citenamefont {Liu}}]{ren-ACSNano2013}%
  \BibitemOpen
  \bibfield  {author} {\bibinfo {author} {\bibfnamefont {M.}~\bibnamefont
  {Ren}}, \bibinfo {author} {\bibfnamefont {J.}~\bibnamefont {Huang}}, \bibinfo
  {author} {\bibfnamefont {H.}~\bibnamefont {Cai}}, \bibinfo {author}
  {\bibfnamefont {J.~M.}\ \bibnamefont {Tsai}}, \bibinfo {author}
  {\bibfnamefont {J.}~\bibnamefont {Zhou}}, \bibinfo {author} {\bibfnamefont
  {Z.}~\bibnamefont {Liu}}, \bibinfo {author} {\bibfnamefont {Z.}~\bibnamefont
  {Suo}},\ and\ \bibinfo {author} {\bibfnamefont {A.-Q.}\ \bibnamefont {Liu}},\
  }\bibfield  {title} {\bibinfo {title} {Nano-optomechanical actuator and
  pull-back instability},\ }\href {https://doi.org/10.1021/nn3056687}
  {\bibfield  {journal} {\bibinfo  {journal} {ACS Nano}\ }\textbf {\bibinfo
  {volume} {7}},\ \bibinfo {pages} {1676} (\bibinfo {year} {2013})}\BibitemShut
  {NoStop}%
\bibitem [{\citenamefont {Efremidis}\ and\ \citenamefont
  {Christodoulides}(2022)}]{efrem-cp2022}%
  \BibitemOpen
  \bibfield  {author} {\bibinfo {author} {\bibfnamefont {N.~K.}\ \bibnamefont
  {Efremidis}}\ and\ \bibinfo {author} {\bibfnamefont {D.~N.}\ \bibnamefont
  {Christodoulides}},\ }\bibfield  {title} {\bibinfo {title} {Thermodynamic
  optical pressures in tight-binding nonlinear multimode photonic systems},\
  }\href {https://doi.org/10.1038/s42005-022-01067-y} {\bibfield  {journal}
  {\bibinfo  {journal} {Communications Physics}\ }\textbf {\bibinfo {volume}
  {5}},\ \bibinfo {pages} {286} (\bibinfo {year} {2022})}\BibitemShut {NoStop}%
\bibitem [{\citenamefont {Ren}\ \emph {et~al.}(2023)\citenamefont {Ren},
  \citenamefont {Pyrialakos}, \citenamefont {Wu}, \citenamefont {Jung},
  \citenamefont {Efremidis}, \citenamefont {Khajavikhan},\ and\ \citenamefont
  {Christodoulides}}]{ren-prl2023}%
  \BibitemOpen
  \bibfield  {author} {\bibinfo {author} {\bibfnamefont {H.}~\bibnamefont
  {Ren}}, \bibinfo {author} {\bibfnamefont {G.~G.}\ \bibnamefont {Pyrialakos}},
  \bibinfo {author} {\bibfnamefont {F.~O.}\ \bibnamefont {Wu}}, \bibinfo
  {author} {\bibfnamefont {P.~S.}\ \bibnamefont {Jung}}, \bibinfo {author}
  {\bibfnamefont {N.~K.}\ \bibnamefont {Efremidis}}, \bibinfo {author}
  {\bibfnamefont {M.}~\bibnamefont {Khajavikhan}},\ and\ \bibinfo {author}
  {\bibfnamefont {D.~N.}\ \bibnamefont {Christodoulides}},\ }\bibfield  {title}
  {\bibinfo {title} {Nature of optical thermodynamic pressure exerted in highly
  multimoded nonlinear systems},\ }\href
  {https://doi.org/10.1103/PhysRevLett.131.193802} {\bibfield  {journal}
  {\bibinfo  {journal} {Phys. Rev. Lett.}\ }\textbf {\bibinfo {volume} {131}},\
  \bibinfo {pages} {193802} (\bibinfo {year} {2023})}\BibitemShut {NoStop}%
\bibitem [{\citenamefont {Ren}\ \emph {et~al.}(2024)\citenamefont {Ren},
  \citenamefont {Pyrialakos}, \citenamefont {Wu}, \citenamefont {Efremidis},
  \citenamefont {Khajavikhan},\ and\ \citenamefont
  {Christodoulides}}]{ren-ol2024}%
  \BibitemOpen
  \bibfield  {author} {\bibinfo {author} {\bibfnamefont {H.}~\bibnamefont
  {Ren}}, \bibinfo {author} {\bibfnamefont {G.~G.}\ \bibnamefont {Pyrialakos}},
  \bibinfo {author} {\bibfnamefont {F.~O.}\ \bibnamefont {Wu}}, \bibinfo
  {author} {\bibfnamefont {N.~K.}\ \bibnamefont {Efremidis}}, \bibinfo {author}
  {\bibfnamefont {M.}~\bibnamefont {Khajavikhan}},\ and\ \bibinfo {author}
  {\bibfnamefont {D.~N.}\ \bibnamefont {Christodoulides}},\ }\bibfield  {title}
  {\bibinfo {title} {Dalton's law of partial optical thermodynamic pressures in
  highly multimoded nonlinear photonic systems},\ }\href
  {https://doi.org/10.1364/OL.517715} {\bibfield  {journal} {\bibinfo
  {journal} {Opt. Lett.}\ }\textbf {\bibinfo {volume} {49}},\ \bibinfo {pages}
  {1802} (\bibinfo {year} {2024})}\BibitemShut {NoStop}%
\bibitem [{\citenamefont {Efremidis}\ and\ \citenamefont
  {Christodoulides}(2024)}]{efrem-ol2024}%
  \BibitemOpen
  \bibfield  {author} {\bibinfo {author} {\bibfnamefont {N.~K.}\ \bibnamefont
  {Efremidis}}\ and\ \bibinfo {author} {\bibfnamefont {D.~N.}\ \bibnamefont
  {Christodoulides}},\ }\bibfield  {title} {\bibinfo {title} {Statistical
  mechanics and pressure of composite multimoded weakly nonlinear optical
  systems},\ }\href {https://doi.org/10.1364/OL.511787} {\bibfield  {journal}
  {\bibinfo  {journal} {Opt. Lett.}\ }\textbf {\bibinfo {volume} {49}},\
  \bibinfo {pages} {2777} (\bibinfo {year} {2024})}\BibitemShut {NoStop}%
\bibitem [{\citenamefont {Haus}\ and\ \citenamefont
  {Kogelnik}(1976)}]{haus-josa1976}%
  \BibitemOpen
  \bibfield  {author} {\bibinfo {author} {\bibfnamefont {H.~A.}\ \bibnamefont
  {Haus}}\ and\ \bibinfo {author} {\bibfnamefont {H.}~\bibnamefont
  {Kogelnik}},\ }\bibfield  {title} {\bibinfo {title} {Electromagnetic momentum
  and momentum flow in dielectric waveguides},\ }\href
  {https://doi.org/10.1364/JOSA.66.000320} {\bibfield  {journal} {\bibinfo
  {journal} {J. Opt. Soc. Am.}\ }\textbf {\bibinfo {volume} {66}},\ \bibinfo
  {pages} {320} (\bibinfo {year} {1976})}\BibitemShut {NoStop}%
\bibitem [{\citenamefont {Pathria}\ and\ \citenamefont
  {Beale}(2011)}]{pathr-2011}%
  \BibitemOpen
  \bibfield  {author} {\bibinfo {author} {\bibfnamefont {R.}~\bibnamefont
  {Pathria}}\ and\ \bibinfo {author} {\bibfnamefont {P.~D.}\ \bibnamefont
  {Beale}},\ }\href@noop {} {\emph {\bibinfo {title} {Statistical
  Mechanics}}},\ \bibinfo {edition} {3rd}\ ed.\ (\bibinfo  {publisher}
  {Butterworth-Heinemann},\ \bibinfo {address} {Oxford},\ \bibinfo {year}
  {2011})\BibitemShut {NoStop}%
\bibitem [{\citenamefont {Efremidis}(2021)}]{efrem-pra2021ssh}%
  \BibitemOpen
  \bibfield  {author} {\bibinfo {author} {\bibfnamefont {N.~K.}\ \bibnamefont
  {Efremidis}},\ }\bibfield  {title} {\bibinfo {title} {Topological photonic
  su-schrieffer-heeger-type coupler},\ }\href
  {https://doi.org/10.1103/PhysRevA.104.053531} {\bibfield  {journal} {\bibinfo
   {journal} {Phys. Rev. A}\ }\textbf {\bibinfo {volume} {104}},\ \bibinfo
  {pages} {053531} (\bibinfo {year} {2021})}\BibitemShut {NoStop}%
\bibitem [{\citenamefont {Okamoto}(2006)}]{okamo-2006}%
  \BibitemOpen
  \bibfield  {author} {\bibinfo {author} {\bibfnamefont {K.}~\bibnamefont
  {Okamoto}},\ }\href@noop {} {\emph {\bibinfo {title} {Fundamentals of optical
  waveguides}}}\ (\bibinfo  {publisher} {Academic Press},\ \bibinfo {address}
  {Burlington},\ \bibinfo {year} {2006})\BibitemShut {NoStop}%
\end{thebibliography}

\newcommand{\noopsort[1]}{} \newcommand{\singleletter}[1]{#1}

\end{document}